\documentclass[prd,preprintnumbers,nofootinbib,superscriptaddress,tightenlines,twocolumn,notitlepage]{revtex4-2}
\usepackage{amssymb}
\usepackage[utf8]{inputenc}
\usepackage[main=english]{babel}
\usepackage{microtype}

\usepackage{graphicx}
\usepackage[usenames,dvipsnames]{xcolor}

\usepackage{diagbox}
\usepackage{booktabs}

\usepackage{amsmath,amsfonts}
\usepackage{mathrsfs}
\usepackage{slashed}
\usepackage{bm}
\usepackage{bbold}

\usepackage{outlines}
\usepackage{enumitem}
\setenumerate[1]{label=\Roman*.}
\setenumerate[2]{label=\Alph*.}
\setenumerate[3]{label=\roman*.}
\setenumerate[4]{label=\alph*.}

\usepackage{titlesec}
\titlespacing{\section}{0pt}{8pt plus 2pt minus 2pt}{5pt plus 1pt minus 1pt}
\titlespacing{\subsection}{0pt}{8pt plus 2pt minus 2pt}{5pt plus 1pt minus 1pt}

\newcommand*\dd{\mathop{}\!\mathrm{d}}
\newcommand{\ddu}{\underline{\dd}}
\newcommand{\sh}[1]{\slashed{#1}}

\def\lf{\left}
\def\rg{\right}
\def\hs{\hspace}

\begin{document}

\title{Pion Electromagnetic Form Factor from Bethe-Salpeter Amplitudes with Appropriate Kinematics}

\author{Shaoyang Jia}
\email[]{syjia@anl.gov}
\affiliation{Physics Division, Argonne National Laboratory, Argonne, IL 60439 USA}

\author{Ian Clo\"{e}t}
\affiliation{Physics Division, Argonne National Laboratory, Argonne, IL 60439 USA}


\begin{abstract}
Within the framework provided by quantum chromodynamics's Schwinger-Dyson equations (SDEs), the pion's electromagnetic form factor is computed using solutions of the Bethe-Salpeter equation (BSE) that have finite spacial momentum, and therefore allow the appropriate kinematics for a given momentum transfer. This removes, for the first time, a limiting approximation in previous SDE calculations where rest-frame solutions to the BSE are used for both the initial and final pion states. In performing these calculations, the rainbow-ladder truncation to the SDEs in the Landau gauge is used, with the quark-gluon interaction given by the Maris-Tandy model. Using Bethe-Salpeter amplitudes (BSAs) that have the correct spacial momentum for a given momentum transfer, $Q^2$, has a dramatic impact on results for the pion's electromagnetic form factor. The difference between results that use rest-frame BSAs and those with correct kinematics is less that 10\% for $0 \leqslant Q^2 \lesssim 1\,$GeV$^2$, however, these corrections grow with increasing momentum transfer and approach 100\% for $Q^2 \simeq 3\,$GeV$^2$. 
\end{abstract}

\maketitle
\section{INTRODUCTION\label{sc:intro}}
The homogeneous Bethe-Salpeter equation (BSE) encodes the structure of a relativistic two-body bound state in terms of its Bethe-Salpeter amplitude (BSA) as its solution. Inputs to such a dynamic equation are the propagators of the constituents and their interactions~\cite{Krassnigg:2009zh,Eichmann:2009zx,Blank:2011qk}. The complexity of the BSE scales quadratically with respect to the number of scalar components of the BSA, which is determined by the types of particles of the bound state and its constituents. In this article, we study the structure of the pion in terms of its valence quarks as a pseudoscalar bound state in quantum chromodynamics (QCD), resulting in $4$ scalar components. 

For a given spacial momentum of the pion, we solve for the BSAs numerically from the BSE in the Euclidean space~\cite{Maris:2005tt}. As one input to the BSE, the propagators of light quarks are determined from their Schwinger-Dyson equation (SDE), which requires the gluon propagator and the quark-gluon $3$-point function as inputs. Another input to the BSE is the quark $4$-point function, the SDE of which involves higher $n$-point functions. As an instance of the rainbow-ladder truncation for these equations, we work in the Landau gauge to apply the Maris-Tandy model~\cite{Maris:1999nt}. 

Within this truncation, we first solve the propagators of the light quarks with spacelike momenta from their SDE. We then project the BSE onto a chosen set of Dirac bases to obtain coupled scalar equations. The mass of the bound state corresponds to the solution with a unit eigenvalue. When sampled by the BSE, quark propagators at complex momenta are computed from their self-energy through its SDE~\cite{Dorkin:2013rsa, Dorkin:2014lxa, Windisch:2016iud, Jia:2024etj}. 

In the rest frame of the bound state, its BSA is applied to determines the decay constant and the vertex residue. The scalar components of this BSA depend on a radial momentum and a temporal angle. Although the BSE in its original formulation is Lorentz invariant, it is practical to choose a frame of reference while working in the Euclidean space. The BSA obtained from the rest frame does not uniquely determine the bound state structure when it acquires a nonzero spacial momentum. The electromagnetic (EM) form factor in the impulse approximation therefore requires solutions of the BSE in boost frames, where non-trivial dependence of the BSA on a spacial angle arises. Keeping track of this dependence is different from the Chebyshev polynomial expansions~\cite{Maris:1999bh,Maris:2005tt}. Solving the BSE while maintaining the Lorentz covariance can be achieved in the Minkowski space~\cite{Jia:2023imt,Jia:2024dbp}.

The Wick rotation is applied to derive the BSE for the scalar functions of the BSA in the Euclidean space. The integral with respect to the loop momentum is represented in $4$-dimensional spherical coordinates. Similar steps are also taken for the normalization of the BSA, the decay constant, the vertex residue, and the EM form factor. These Euclidean-space expressions are evaluated numerically after choosing a set of grid points for variables of the scalar functions. With a typical infrared (IR) scale of the Maris-Tandy model and a given light-quark mass, we solve the rest-frame BSE for the pion. This quark mass is adjusted such that the bound state mass becomes the physical pion mass. In the meantime, the strength of the IR term is adjusted to reproduce the experimental decay constant. 

After choosing the model parameters, we solve the BSE in boost frames and subsequently the EM form factor with spacelike photon. The EM form factor is computed with two BSAs each carrying half of the photon momentum. Particularly we base our work on Refs.~\cite{Maris:2005tt,Maris:1999bh,Maris:2000sk} with the improvement of solving the BSE for a given discretized grid of the momentum variables. The explicit expressions in full detail at the component level used by the numerical implementation of both the BSE and the EM form factor are also given. We provide the first solutions of the BSE with the discretized grid for momentum variables in both the rest frame and the boost frame of the pion. To illustrate the need of boost-frame BSAs when the photon momentum is large, this form factor is compared to the one from the rest-frame BSA. Another input to the EM form factor is the quark-photon vertex, which could be solved from its inhomogeneous BSE~\cite{Maris:1999bh}. Instead of solving this dynamic equation, we apply the Ball-Chiu vertex in conjunction with an Ansatz transverse term~\cite{Maris:1999bh}. While the EM form factor in the timelike region~\cite{Miramontes:2022uyi,Sauli:2022ild} and other transitions of the pion are also accessible via these BSAs~\cite{Maris:2000sk, Maris:2002mz,Jarecke:2002xd,Eichmann:2017wil,Weil:2017knt,Weil:2017moo}, they are beyond the scope of this article.

This article is organized as follows. Section~\ref{sc:intro} is the introduction. Section~\ref{sc:BSE} derives the decomposition of the BSE into scalar components and the resulting Euclidean-space expressions. The normalization, the decay constant, and the vertex residue in terms of the BSA are given in Section~\ref{sc:static}. Section~\ref{sc:emff} contains the EM form factor in the impulse approximation. Numerical settings, model parameters, and observables of the pion are presented in Sect.~\ref{sc:results}. Section~\ref{sc:summary} is the summary. Supporting derivations are given in the Appendices. 
\section{BETHE-SALPETER EQUATION\label{sc:BSE}}
\subsection{BSE in Minkowski Space}
\begin{figure*}[tbp]
	\centering\includegraphics[width=0.65\linewidth]{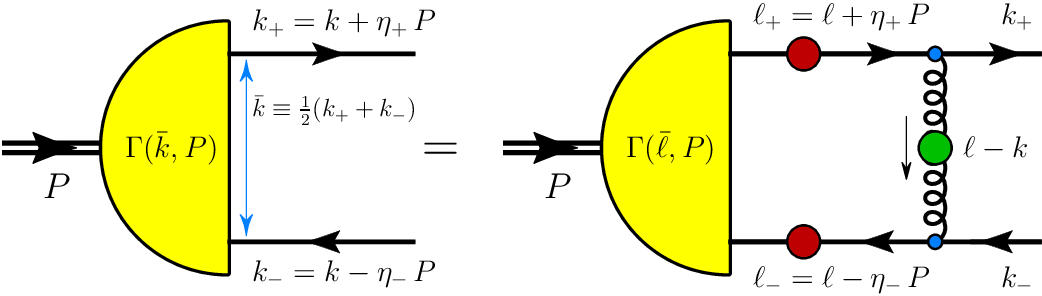}
	\caption{BSE for the pseudoscalar bound states of a fermion and an antifermion in the rainbow-ladder truncation. Here the yellow blobs represent the BSAs of the bound state. While the red blobs and the green blob stand for the dressed propagators of the quarks and the gluons. The momenta $P$, $k_\pm$, and $q$ are assigned to the bound state, the constituent quarks, and the gluon, respectively.}
	\label{fig:psbsefermions}
\end{figure*}

In rainbow-ladder truncation, the homogeneous BSE is illustrated in Fig.~\ref{fig:psbsefermions} and in Minkowski space reads~\cite{Krassnigg:2009zh,Eichmann:2009zx,Blank:2011qk}:
\begin{equation}
	\Gamma(k,P) = ig^2C_{\mathrm{F}}\!\int\! \ddu \ell\,
	\gamma^\mu S(\ell_+)\Gamma(\ell,P)S(\ell_-)\gamma^\nu
	\Delta_{\mu\nu}(\ell-k), 
	\label{eq:ps_BSE}
\end{equation}
where $g$ is the elementary charge of the strong interaction, $C_{\mathrm{F}} = 4/3$ is a color factor corresponding to number of colors ${N_{\mathrm{c}}=3}$, the internal quark momenta are given by $\ell_\pm = \ell \pm \eta_\pm P$ with $\eta_+ + \eta_- = 1$, and in $4$-dimensional spacetime the integral measure is $\int \ddu\ell \equiv \int_{-\infty}^{+\infty}\dd^4\ell / (2\pi)^4$. The dressed gluon propagator  in the Landau gauge is given by
\begin{equation}
	\Delta_{\mu\nu}(q) = \left(-g^{\mu\nu} + q^\mu q^\nu / q^2 \right)\mathcal{G}(q^2),
\end{equation}
where in this work the scalar function $\mathcal{G}(q^2)$ is specified in Euclidean space by the Maris-Tandy model~\cite{Maris:1999nt}. The dressed-quark propagator has the general form
\begin{equation}
	S(k) = \sh{k}\,\sigma_{\mathrm{v}}(k^2) + \sigma_{\mathrm{s}}(k^2) = \lf[\sh{k}\,A(k^2) + B(k^2)\rg]^{-1},
	\label{eq:dec_SF}
\end{equation} 
where $\sigma_{\mathrm{v}}(k^2) = A(k^2)/D(k^2)$, $\sigma_{\mathrm{s}}(k^2) = -B(k^2) / D(k^2)$, the denominator function reads $D(k^2) = A^2(k^2)\,k^2- B^2(k^2)$, and the mass function is given by $\mathcal{M}(k^2) = -B(k^2) / A(k^2)$. These scalar functions are obtained by solving the gap equation for the quark propgator~\cite{Williams:2007zzh,Dorkin:2013rsa,Dorkin:2014lxa, Windisch:2016iud,Jia:2024etj}. The general form of the BSA for a pseudoscalar bound state reads
\begin{equation}
	\Gamma(k,P) = \gamma_5\,T(k,P)\,\mathbb{G}(k^2,k\cdot P), 
	\label{eq:ps_BSA_matrix}
\end{equation}
where we have defined the vectors
\begin{align}
	\label{eq:def_Tj_basis}
	T(k,P) &= 
	\begin{pmatrix}
		\mathbb{1} & \slashed{P} & \slashed{k} & -i\sigma_{kP}
	\end{pmatrix}, \\
	\label{eq:def_bbm_G_bsa}
	\mathbb{G}(k^2,k\cdot P) &= 
	\begin{pmatrix}
		E(k^2,k\cdot P) \\
		F(k^2,k\cdot P) \\
		G(k^2,k\cdot P) \\
		H(k^2,k\cdot P)
	\end{pmatrix},
\end{align}
and $\sigma_{kP} \equiv k^\mu P^\nu\sigma_{\mu\nu}$, $\sigma^{\mu\nu} = i(\gamma^\mu\gamma^\nu -\gamma^\nu\gamma^\mu)/2$. Here $k$ is the relative momentum of the constituents, and $P$ is total the momentum of the bound state. The quark momenta are related to $k$ and $P$ by
\begin{equation}
	k_\pm = k \pm \eta_\pm P, 
	\label{eq:def_kpm}
\end{equation}
such that $P = k_+ - k_-$, $k = \eta_- k_+ + \eta_+ k_-$, and a typical choice momentum partitioing parameter is $\eta_\pm = m_\pm / (m_+ + m_-)$ where the masses of the constituents are $m_+$ and $m_-$. This results in $\eta_{\pm} = 0.5$ for the pion in the isospin symmetric limit. 

Following the procedure given in the Appendices, the homogeneous BSE of Eq.~\eqref{eq:ps_BSE} becomes a set of coupled integral equations for the scalar functions in Eq.~\eqref{eq:def_bbm_G_bsa}, specified by a $4$-by-$4$ matrix of integral operators. To derive these operators, the first step is to expand the Bethe-Salpeter (BS) wave function defined as $S(\ell_+)\,\Gamma(\ell,P)\,S(\ell_-)$ in terms of the Dirac bases similar to Eq.~\eqref{eq:ps_BSA_matrix}. Trace projections are then taken on both sides of the BSE giving
\begin{align}
	&\mathbb{G}(k^2,k\cdot P) = - ig^2 C_{\mathrm{F}} \int \ddu \ell\ 
	\mathcal{G}[(\ell-k)^2] \nonumber\\
	& \times \hat{\mathbb{T}}(k^2,k\cdot P, \ell^2, \ell\cdot P, k\cdot \ell)\,
	\mathbb{M}(\ell^2,\ell\cdot P)\,
	\mathbb{G}(\ell^2,\ell\cdot P).
	\label{eq:ps_BSE_projected}
\end{align}
Matrices $\mathbb{M}(\ell^2,\ell\cdot P)$ and $\hat{\mathbb{T}}(k^2,k\cdot P, \ell^2, \ell\cdot P, k\cdot \ell)$ are defined in Eq.~\eqref{eq:def_matrix_bbM} and Eq.~\eqref{eq:def_That_bbm}, respectively. Solving Eq.~\eqref{eq:ps_BSE_projected} directly in the Minkowski space requires the analytic properties of the BSA~\cite{Kusaka:1995za,Kusaka:1997xd,Jia:2023imt}. As such, we follow the usual alternative in the SDE framework and solve Eq.~\eqref{eq:ps_BSE_projected} numerically in Euclidean space after specifying a reference frame.
\subsection{BSE in Euclidean Space -- at rest}
We would like to formulate Eq.~\eqref{eq:ps_BSE_projected} such that functions in Eq.~\eqref{eq:def_bbm_G_bsa} are only needed with spacelike $k^\mu$. The metric for a Minkowski-space inner product is $g^{\mu\nu} = \mathrm{diag}\{1,\,-1,\,-1,\,-1\}$, while the Euclidean space metric is $\delta^{ij} = \{1,\,1,\,1,\,1\}$. Denoting functions and variables in the Euclidean space by the subscript ``$\mathrm{E}$'', the Euclidean-space counterpart of $k^\mu$ becomes $k_{\mathrm{E}}^j$. Its components are specified by the Wick rotation as $k^0 = ik_{\mathrm{E}}^4$ and $k^j = k_{\mathrm{E}}^j$ for $j\in\{1,\,2,\,3\}$. In terms of hyperspherical coordinates, the loop momentum in Eq.~\eqref{eq:ps_BSE_projected} becomes
\begin{align}
	\ell^j_{\mathrm{E}} & = \ell_{\mathrm{E}} \, (\sin\psi\sin\theta\sin\phi, \nonumber \\ 
	&\hs*{15mm}
	\sin\psi\sin\theta\cos\phi,\ \sin\psi\cos\theta,\ \cos\psi), 
	\label{eq:def_loop_momentum}
\end{align}
with $\ell_{\mathrm{E}} = |\ell_{\mathrm{E}}|$ the modulus of $\ell^j_{\mathrm{E}}$, the temporal angle $\psi\in [0,\pi]$, the spacial angle $\theta\in[0,\pi]$, and the azimuthal angle $\phi\in[0,2\pi]$. 
The integration measure then becomes
\begin{align}
	\int \ddu \ell & = 
	\dfrac{i}{(2\pi)^4} \int_0^\infty \dd \ell_{\mathrm{E}}\, \ell^3_{\mathrm{E}} \nonumber\\
	& \quad \times \int_{-1}^{1}\dd z'\ \sqrt{1-z'^2} \int_{-1}^{1}\dd y' \int_{0}^{2\pi}\dd \phi, \label{eq:itg_measure_dkp_ori}
\end{align}
where we have defined $y' = \cos \theta$ and $z' = \cos \psi$. A choice of reference frame is needed in order to convert Eq.~\eqref{eq:ps_BSE_projected} into integral equations for scalar functions. 

In the rest frame of the bound state, momenta of the BSA in Eq.~\eqref{eq:ps_BSA_matrix} become
\begin{subequations}
	\label{eq:P_k_Euc_hyperspherical}
	\begin{align}
		P_{\mathrm{E}}^j & = (0,\, 0,\, 0,\, -iM), \label{eq:P_Euc_hyperspherical} \\
		k_{\mathrm{E}}^j & = (0,\, 0,\, k_{\mathrm{E}}\sqrt{1-z^2},\, z\,k_{\mathrm{E}}),
	\end{align}
\end{subequations}
with $M$ the bound state mass. For $\ell^j_{\mathrm{E}}$ in Eq.~\eqref{eq:def_loop_momentum}, we have
${k_{\mathrm{E}} \cdot \ell_{\mathrm{E}}} = |k_{\mathrm{E}}|\, |\ell_{\mathrm{E}}|\,\zeta(z,z',y')$ with
\begin{equation}
	\zeta(z,z',y') = zz' + \sqrt{1-z^2}\,\sqrt{1-z'^2}\,y'. 
	\label{eq:def_zeta_rest_frame}
\end{equation} 
The momenta for the quark propagators are given by
\begin{equation}
	\ell^2_{\pm\mathrm{E}} = \ell^2_{\mathrm{E}} - \eta^2_\pm M^2 \mp 2i\eta_\pm \ell_{\mathrm{E}}Mz' , \label{eq:quark_Euc_momentum_rest_frame} 
\end{equation}
which fall within the region of the complex-momentum plane enclosed by a parabola. Let us define the Schwinger functions of the propagators in the Euclidean space as $A_{\mathrm{E}}(k^2_\mathrm{E}) = A(-k^2_\mathrm{E})$, $B_{\mathrm{E}}(k^2_{\mathrm{E}}) = - B(-k^2_{\mathrm{E}})$, $D_{\mathrm{E}}(k^2_\mathrm{E}) = -D(-k^2_\mathrm{E})$, ${ \mathcal{M}_{\mathrm{E}}(k^2_{\mathrm{E}}) = B_{\mathrm{E}}(k^2_{\mathrm{E}})/A_{\mathrm{E}}(k^2_{\mathrm{E}}) }$, and $\mathcal{G}_{\mathrm{E}}(q_{\mathrm{E}}^2) = - \mathcal{G}(-q_{\mathrm{E}}^2)$. We also define the Euclidean-space scalar functions of the BSA from their Minkowski-space counterparts by
$\mathbb{G}_\mathrm{E}(k_\mathrm{E}^2, z) = \mathbb{G}(-k^2_{\mathrm{E}}, ik_{\mathrm{E}}Mz)$ such that
\begin{equation}
	\mathbb{G}_{\mathrm{E}}(k_\mathrm{E}^2, z)=
	\begin{pmatrix}
		E_{\mathrm{E}}(k_\mathrm{E}^2, z) \\
		F_{\mathrm{E}}(k_\mathrm{E}^2, z) \\
		G_{\mathrm{E}}(k_\mathrm{E}^2, z) \\
		H_{\mathrm{E}}(k_\mathrm{E}^2, z)
	\end{pmatrix}, \label{eq:def_bbm_GE_bsa}
\end{equation}
While the trace matrix in Eq.~\eqref{eq:ps_BSE_projected} becomes
\begin{align}
	& \hat{\mathbb{T}}_{\mathrm{E}}(k_{\mathrm{E}}^2, z, \ell^2_{\mathrm{E}}, z',y') \nonumber\\
	& \hspace{-0.3cm} = \hat{\mathbb{T}} \big( -k^2_{\mathrm{E}}, ik_{\mathrm{E}}Mz, -\ell^2_{\mathrm{E}}, i\ell_{\mathrm{E}}Mz', -k_{\mathrm{E}}\ell_{\mathrm{E}} \,\zeta(z,z',y') \big) . \label{eq:def_TE_rest_frame}
\end{align}
Combining Eq.~\eqref{eq:def_That_bbm} and Eq.~\eqref{eq:def_TE_rest_frame}, we obtain Eq.~\eqref{eq:Euclidean_bbm_trace_matrix} as the explicit expression for $\hat{\mathbb{T}}_{\mathrm{E}}{(k_{\mathrm{E}}^2, z, \ell^2_{\mathrm{E}}, z',y')}$, where the factor of $(1-z^2)^{-1/2}$ does not contribute to any end-point singularities because the $y'$ dependence in the gluon propagator always enters as $(1-z^2)^{1/2}y'$. 

We then represent $\mathbb{M}(\ell^2,\ell\cdot P)$ in Eq.~\eqref{eq:ps_BSE_projected} in the rest frame by ${\mathbb{M}_{\mathrm{E}}^\pm(\ell^2_{\mathrm{E}},z')}$ defined as
\begin{subequations} \label{eq:def_bbm_pm_euc}
	\begin{align}
		&\mathbb{M}_{\mathrm{E}}^+(\ell^2_{\mathrm{E}},z') = -\sigma^+_{s\mathrm{E}}(\ell^2_{+\mathrm{E}}) \mathbb{1}_{4} + \sigma^+_{v\mathrm{E}}(\ell^2_{+\mathrm{E}}) \nonumber \\
		& \hspace{0.5cm}\times
		\left[ k_{\mathrm{L}}(-\ell^2_{\mathrm{E}}, i\ell_{\mathrm{E}}Mz') 
		+ \eta_{+} P_{\mathrm{L}}(i\ell_{\mathrm{E}}Mz', M^2) \right],
		\label{eq:def_bbmm_pm_euc} \\
		& \mathbb{M}_{\mathrm{E}}^-(\ell^2_{\mathrm{E}},z') = \sigma^-_{s\mathrm{E}}(\ell^2_{-\mathrm{E}})\, \mathbb{1}_{4} + \sigma^-_{v\mathrm{E}}(\ell^2_{-\mathrm{E}}) \nonumber \\
		& \hspace{0.5cm}\times 
		\left[ k_{\mathrm{R}}(-\ell^2_{\mathrm{E}}, i\ell_{\mathrm{E}}Mz') - \eta_{-} P_{\mathrm{R}}(i\ell_{\mathrm{E}}Mz', M^2) \right],
		\label{eq:def_bbmp_pm_euc}
	\end{align} 
\end{subequations}
with $\ell^2_{\pm\mathrm{E}}$ given by Eq.~\eqref{eq:quark_Euc_momentum_rest_frame}. Here $\sigma^{\pm}_{\mathrm{v}/\mathrm{s}\,\mathrm{E}}(k^2_{\mathrm{E}})$ are
defined as Euclidean versions of those in Eq.~\eqref{eq:dec_SF} by
\begin{subequations}\label{eq:Euc_quark_prop_functions}
	\begin{align}
		\sigma^{\pm}_{v\mathrm{E}}(k^2_{\mathrm{E}}) & = A^{\pm}_{\mathrm{E}}(k^2_{\mathrm{E}}) / D^{\pm}_{\mathrm{E}}(k^2_{\mathrm{E}}), \\
		\sigma^{\pm}_{s\mathrm{E}}(k^2_{\mathrm{E}}) & = B^{\pm}_{\mathrm{E}}(k^2_{\mathrm{E}}) / D^{\pm}_{\mathrm{E}}(k^2_{\mathrm{E}}) ,
	\end{align} 
\end{subequations}
with superscripts specifying quark flavors. 
They stand for the scalar components of the valence anti-quark in Eq.~\eqref{eq:def_bbmm_pm_euc}. While those in Eq.~\eqref{eq:def_bbmp_pm_euc} correspond to the valence quark. The BSE in Eq.~\eqref{eq:ps_BSE_projected} then becomes
\begin{align}
	\mathbb{G}_{\mathrm{E}}(k^2_{\mathrm{E}}, z) &= -\frac{g^2\,C_{\mathrm{F}}}{8\pi^3} 
	\int_{0}^{\infty}\dd \ell_{\mathrm{E}}\, \ell^3_{\mathrm{E}}
	\int_{-1}^{1}\dd z' \sqrt{1-z'^2} \int_{-1}^{1}\dd y' \nonumber\\
	&\quad \times  
	\mathcal{G}_{\mathrm{E}}\left(\left(\ell_{\mathrm{E}}-k_{\mathrm{E}}\right)^2\right)
	\hat{\mathbb{T}}_{\mathrm{E}}(k_{\mathrm{E}}^2, z, \ell^2_{\mathrm{E}}, z',y') \nonumber\\
	&\quad \times
	\mathbb{M}_{\mathrm{E}}^+(\ell^2_{\mathrm{E}},z')\,
	\mathbb{M}_{\mathrm{E}}^-(\ell^2_{\mathrm{E}},z')\,
	\mathbb{G}_{\mathrm{E}}(\ell^2_{\mathrm{E}}, z'),
	\label{eq:ps_BSE_Euc_rest_frame}
\end{align}
in the rest frame of the bound state. Because the $y'$ integral only involves the gluon propagator and ${ \hat{\mathbb{T}}_{\mathrm{E}}(k^2_{\mathrm{E}},z,\ell^2_{\mathrm{E}},z',y') }$, it samples known functions when applying the Maris-Tandy model.
\subsection{BSE in Euclidean Space -- finite spacial momentum}
When the bound state carries a spacial momentum of $Q/2$, let us define an angle $\varphi$ through $\sinh\varphi = Q/{(2M)}$, 
which is different from the azimuthal angle $\phi$ in Eq.~\eqref{eq:def_loop_momentum}. We then choose the reference frame by aligning the spacial momentum of the bound state to the $z$-axis such that
\begin{equation}
	P_{\mathrm{E}}^j = 
	\begin{pmatrix}
		0 & 0 & M\,\sinh\varphi & -iM\,\cosh\varphi
	\end{pmatrix}. \label{eq:def_bound_state_momentum_boosted_frame}
\end{equation}
The internal momentum of the BSA in Eq.~\eqref{eq:ps_BSA_matrix} becomes
\begin{equation}
	k_{\mathrm{E}}^j = 
	\begin{pmatrix}
		0 & k_{\mathrm{E}}\sqrt{(1-y^2)(1-z^2)} & k_{\mathrm{E}}y\sqrt{1-z^2} & k_{\mathrm{E}}z
	\end{pmatrix} .
\end{equation}
While the loop momentum $\ell^j_{\mathrm{E}}$ in Eq.~\eqref{eq:ps_BSE_projected} is given by Eq.~\eqref{eq:def_loop_momentum}. Within this frame, the momenta of the quark propagators are given by
\begin{equation}
	\ell^2_{\pm\mathrm{E}} = \ell^2_{\mathrm{E}} - \eta^2_\pm M^2 
	\pm 2\eta_\pm \ell_{\mathrm{E}}\cdot P_{\mathrm{E}}, 
	\label{eq:quark_Euc_momentum_boosted_frame}
\end{equation}
which locate within a parabolic region in the complex-momentum plane. 

The scalar functions of the BSA in the Euclidean space are related to those in the Minkowski space by
\begin{equation}
	\mathbb{G}_\mathrm{E}(k_\mathrm{E}^2, y, z) = \mathbb{G}(-k^2_{\mathrm{E}}, -k_{\mathrm{E}}\cdot P_{\mathrm{E}}). \label{eq:def_bbm_G_E_boosted_frame}
\end{equation}
Notice that the dependence on the variable $y$ enters in addition to $k_\mathrm{E}^2$ and $z$ that are present in the rest-frame expressions. Applying procedures similar to those in the previous Subsection, the BSE in this frame becomes
\begin{align}
	& \mathbb{G}_{\mathrm{E}}(k^2_{\mathrm{E}}, y, z) = -\frac{g^2\,C_{\mathrm{F}}}{(2\pi)^4}\int_{0}^{\infty}\dd \ell_{\mathrm{E}}\, \ell^3_{\mathrm{E}} 
	\int_{-1}^{+1}\dd z'\sqrt{1-z'^2} 
	\nonumber\\
	& \times \! \int_{-1}^{1}\dd y'\int_{0}^{2\pi}\dd\phi\, \mathcal{G}_{\mathrm{E}}((\ell_{\mathrm{E}}-k_{\mathrm{E}})^2)
	\hat{\mathbb{T}}_{\mathrm{E}}(k_{\mathrm{E}}^2,y,z,\ell^2_{\mathrm{E}},y',z',\phi)\nonumber\\
	& \times 
	\mathbb{M}_{\mathrm{E}}^+(\ell^2_{\mathrm{E}}, y', z')\,
	\mathbb{M}_{\mathrm{E}}^-(\ell^2_{\mathrm{E}}, y', z')\,
	\mathbb{G}_{\mathrm{E}}(\ell^2_{\mathrm{E}}, y', z').  
	\label{eq:ps_BSE_Euc_boosted_frame}
\end{align}
Here through Eq.~\eqref{eq:def_That_bbm} we have defined 
\begin{align}
	& \hat{\mathbb{T}}_{\mathrm{E}}(k^2_{\mathrm{E}},y,z,\ell^2_{\mathrm{E}},y',z',\phi) \nonumber\\
	& = 
	\hat{\mathbb{T}}(-k^2_{\mathrm{E}},-k_{\mathrm{E}}\cdot P_{\mathrm{E}}, -\ell^2_{\mathrm{E}}, -\ell_{\mathrm{E}}\cdot P_{\mathrm{E}}, -k_{\mathrm{E}}\cdot \ell_{\mathrm{E}}) , \label{eq:def_TE_boosted_frame}
\end{align}
which is explicitly given as Eq.~\eqref{eq:Euclidean_bbm_trace_matrix_boosted_frame}. While the mass matrices in this frame are defined as
\begin{subequations} \label{eq:def_bbm_pm_euc_boosted_frame}
	\begin{align}
		& \mathbb{M}^+_{\mathrm{E}}(\ell^2_{\mathrm{E}},y',z') = -\sigma^+_{\mathrm{sE}}(\ell^2_{+\mathrm{E}}) \mathbb{1}_{4} + \sigma^+_{\mathrm{vE}}(\ell^2_{+\mathrm{E}}) \nonumber\\
		& \times \left[ k_{\mathrm{L}}(-\ell^2_{\mathrm{E}}, - \ell_{\mathrm{E}}\cdot P_{\mathrm{E}}) 
		+ \eta_{+}P_{\mathrm{L}}(-\ell_{\mathrm{E}}\cdot P_{\mathrm{E}}, M^2 ) \right], \\
		& \mathbb{M}^{-}_{\mathrm{E}}(\ell^2_{\mathrm{E}},y',z') = \sigma^-_{\mathrm{sE}}(\ell^2_{-\mathrm{E}})\mathbb{1}_{4} + \sigma^-_{\mathrm{vE}}(\ell^2_{-\mathrm{E}}) \nonumber\\
		& \times \left[ k_{\mathrm{R}}(-\ell^2_{\mathrm{E}}, -\ell_{\mathrm{E}}\cdot P_{\mathrm{E}})
		- \eta_{-}P_{\mathrm{R}}(-\ell_{\mathrm{E}}\cdot P_{\mathrm{E}}, M^2 ) \right],
	\end{align}
\end{subequations}
with $\ell^2_{\pm\mathrm{E}}$ given by Eq.~\eqref{eq:quark_Euc_momentum_boosted_frame}. Functions $\sigma^{\pm}_{v\mathrm{E}}(k^2_{\mathrm{E}})$ and $\sigma^{\pm}_{s\mathrm{E}}(k^2_{\mathrm{E}})$ in Eq.~\eqref{eq:def_bbm_pm_euc_boosted_frame} are defined by Eq.~\eqref{eq:Euc_quark_prop_functions}. These functions for the quark and for the anti-quark are the same when modeling the pion in the isospin symmetric limit. 

Equation~\eqref{eq:ps_BSE_Euc_boosted_frame} is the BSE when the spacial momentum of the bound state can be nonzero. In the limit where this momentum becomes zero, Eq.~\eqref{eq:ps_BSE_Euc_boosted_frame} is reduced to Eq.~\eqref{eq:ps_BSE_Euc_rest_frame}. With the gluon propagator specified by the Maris-Tandy model, the integral over the azimuthal angle $\phi$ only involves known functions. 
\section{BSA NORMALIZATION AND STATIC OBSERVABLES\label{sc:static}}
\subsection{BSA Normalization in Minkowski space}
The BSE is homogeneous, leaving the normalization of its solutions unspecified. Within our truncation, the quark $4$-point kernel function is independent of the bound state momentum. The normalization of the pseudoscalar BSA is therefore determined from~\cite{Maris:1999nt,Williams:2009wx}
\begin{align}
	1 & = -6i \lim\limits_{Q^2 \rightarrow P^2} \frac{\partial}{\partial\,Q^2} \int \ddu \ell \nonumber \\
	& \quad \times \mathrm{Tr}\lf[\overline{\Gamma}(\ell,-P)\,S(\ell_+)\,\Gamma(\ell,P)\,S(\ell_-)\rg], \label{eq:BSA_normalization}
\end{align}
where the quark momenta are defined as $\ell_\pm = \ell \pm { \eta_\pm Q }$. To rewrite Eq.~\eqref{eq:BSA_normalization} in scalar functions, let us first apply Eq.~\eqref{eq:ps_BSA_matrix} such that the conjugate BSA defined as $\overline{\Gamma}(\ell,-P) = C\,\Gamma^{\mathrm{T}}(-\ell,-P)\,C^{-1}$ becomes
\begin{equation}
	\overline{\Gamma}(\ell,-P) \\
	= \overline{\mathbb{G}}(\ell^2, \ell\cdot P)
	\overline{T}(\ell, P)
	\, \gamma_5 ,
	\label{eq:Gamma_bar_neg_P}
\end{equation}
with charge conjugate matrix given by ${C = i\gamma^2\gamma^0}$. Here we have defined $\overline{\mathbb{G}}(\ell^2, \ell\cdot P)$ according to
\begin{align}
	\overline{\mathbb{G}}(\ell^2, \ell\cdot P) & = \big[ E(\ell^2,\ell\cdot P), F(\ell^2, \ell\cdot P), \nonumber \\
	& \quad \quad
	G(\ell^2, \ell\cdot P), -H( \ell^2, \ell\cdot P) \big]. 
	\label{eq:def_bbm_G_bar}
\end{align}
and 
\begin{equation}
	\overline{T}(\ell,P) = \left( \mathbb{1},\, \slashed{P},\, \slashed{\ell},\, -i\sigma_{\ell P} \right)^{\mathrm{T}} \label{eq:def_Tj_basis_transpose}
\end{equation}
with the superscript ``T'' standing for the matrix transpose. Because $Q\neq P$ in the factor  $S(\ell_+)\,\Gamma(\ell,P)\,S(\ell_-)$ of Eq.~\eqref{eq:BSA_normalization}, the multiplications of the quark propagators expand the Dirac bases from Eq.~\eqref{eq:def_Tj_basis} to Eq.~\eqref{eq:extended_basis}. Relevant terms in Eq.~\eqref{eq:BSA_normalization} then become
\begin{align}
	& \dfrac{1}{4}\mathrm{Tr}\,\overline{\Gamma}(\ell,-P)\,S(\ell_+)\,\Gamma(\ell,P)\,S(\ell_-) = \overline{\mathbb{G}}(\ell^2,\ell\cdot P)  \nonumber\\
	& \quad \times \mathbb{U}(\ell^2,\ell\cdot P, \ell\cdot Q)\, \mathbb{S}^{+}(\ell^2, \ell\cdot P, \ell\cdot Q, P\cdot Q, Q^2)\nonumber\\
	& \quad \times \mathbb{S}^{-}(\ell^2, \ell\cdot P, \ell\cdot Q, P\cdot Q, Q^2)\,
	\mathbb{G}(\ell^2, \ell\cdot P). \label{eq:trace_normalization}
\end{align}
Here the mapping due to the quark propagators are represented by the following matrices
\begin{subequations}
	\label{eq:prop_matrix_normalization}
	\begin{align}
		& \mathbb{S}^{+}(\ell^2, \ell\cdot P, \ell\cdot Q, P\cdot Q, Q^2)  = \big[ -K_{\mathrm{L}}(\ell^2, \ell\cdot P, \ell\cdot Q) \nonumber\\
		& -\eta_+ Q_{\mathrm{L}}(\ell\cdot Q, P\cdot Q, Q^2) \big] \,\sigma^+_{\mathrm{v}}(\ell^2_+) + I_{\mathrm{L}}\,\sigma^+_{\mathrm{s}}(\ell^2_+), \\
		& \mathbb{S}^{-}(\ell^2, \ell\cdot P, \ell\cdot Q, P\cdot Q, Q^2)  = \big[ K_{\mathrm{R}}(\ell^2, \ell\cdot P) \nonumber\\
		& - \eta_- Q_{\mathrm{R}}(\ell\cdot P, P\cdot Q) \big]\, \sigma^-_{\mathrm{v}}(\ell^2_-) + I_{\mathrm{R}}\,\sigma^-_{\mathrm{s}}(\ell^2_-) ,
	\end{align}
\end{subequations}
with $K_{\mathrm{L/R}}$, $Q_{\mathrm{L/R}}$, and $I_{\mathrm{L/R}}$ given by Eq.~\eqref{eq:def_matrices_normalization}. 
The scalar components of quark propagators are defined according to Eq.~\eqref{eq:dec_SF}. The trace operations in Eq.~\eqref{eq:BSA_normalization} are subsequently reduced into those on the bases. We have consequently defined the matrix ${\mathbb{U}(\ell^2, \ell\cdot P, \ell\cdot Q)}$ in Eq.~\eqref{eq:trace_normalization} according to
\begin{equation}
	\mathbb{U}_{ij}(\ell^2, \ell\cdot P, \ell\cdot Q) = \dfrac{1}{4}\mathrm{Tr} \,T_i(\ell,P)\,T_j(\ell,P,Q), 
\end{equation}
with ${T_i(k,P)}$ and ${T_j(k,P,Q)}$ given by Eq.~\eqref{eq:def_Tj_basis} and Eq.~\eqref{eq:extended_basis}. Nonzero elements of $\mathbb{U}(\ell^2, \ell\cdot P, \ell\cdot Q)$ are $\mathbb{U}_{11} = 1$, $\mathbb{U}_{22} = P^2$, $\mathbb{U}_{23} = \mathbb{U}_{32} = {\ell\cdot P}$, $\mathbb{U}_{25} = {P\cdot Q}$, $\mathbb{U}_{33} = \ell^2$, $\mathbb{U}_{35} = {\ell\cdot Q}$, $\mathbb{U}_{44} = {(\ell\cdot P)^2} - \ell^2P^2$, $\mathbb{U}_{46} = {\ell\cdot QP^2} -{(\ell\cdot P)}{(P\cdot Q)}$, and $\mathbb{U}_{47} = {(\ell\cdot P)}{(\ell\cdot Q)} - {\ell^2}{P\cdot Q}$. The normalization condition then becomes 
\begin{align}
	1 = & -24i \lim\limits_{Q^2\rightarrow P^2}\frac{\partial}{\partial\,Q^2}
	\int \ddu \ell\ \overline{\mathbb{G}}(\ell^2, \ell\cdot P)\, \mathbb{U}(\ell^2, \ell\cdot P, \ell\cdot Q)\nonumber\\
	& \times \mathbb{S}^+(\ell^2, \ell\cdot P, \ell\cdot Q, P\cdot Q, Q^2) \nonumber\\
	& \times \mathbb{S}^-(\ell^2, \ell\cdot P, \ell\cdot Q, P\cdot Q, Q^2)\
	\mathbb{G}(\ell^2, \ell\cdot P). 
	\label{eq:BSA_normalization_components}
\end{align}

\subsection{BSA Normalization in Euclidean space}
In the rest frame of the bound state, we have $P^j_{\mathrm{E}}$ given by Eq.~\eqref{eq:P_Euc_hyperspherical} and $Q^j_{\mathrm{E}} = {(0,\,0,\,0,\,-iQ)}$ in Eq.~\eqref{eq:BSA_normalization_components}. The loop momentum is similar to that in Eq.~\eqref{eq:def_loop_momentum}. We then obtain the following inner products: ${\ell^2 = -\ell^2_{\mathrm{E}}}$, ${P^2 = M^2}$, ${\ell\cdot P = iM\ell_{\mathrm{E}}z}$, ${\ell\cdot Q = iQ\ell_{\mathrm{E}}z}$, and ${ P\cdot Q = MQ }$, with $z = \cos \psi$ representing the temporal angle of $\ell^j_{\mathrm{E}}$. The propagator matrices in Eq.~\eqref{eq:prop_matrix_normalization} in the rest frame are defined as
$ \mathbb{S}^\pm_{\mathrm{E}}{(\ell_{\mathrm{E}}, z, Q)} = - \mathbb{S}^\pm{(-\ell^2_{\mathrm{E}},iM\ell_{\mathrm{E}}z, iQ\ell_{\mathrm{E}}z, QM, Q^2)}$, 
which yield
\begin{subequations}\label{eq:def_Euc_S_pm}
	\begin{align}
		& \mathbb{S}^+_{\mathrm{E}}(\ell_{\mathrm{E}}, z, Q) = \sigma^+_{\mathrm{vE}}(\ell^2_{+\mathrm{E}})\,\big[ -K_{\mathrm{L}}( -\ell^2_{\mathrm{E}}, iM\ell_{\mathrm{E}}z, iQ\ell_{\mathrm{E}}z) \nonumber\\
		& -\eta_+Q_{\mathrm{L}}( iQ\ell_{\mathrm{E}}z, MQ, Q^2 )\big]  + \sigma^+_{\mathrm{sE}}(\ell^2_{+\mathrm{E}})\,I_{\mathrm{L}}, \\
		& \mathbb{S}^-_{\mathrm{E}}(\ell_{\mathrm{E}}, z, Q) = \sigma^-_{\mathrm{vE}}(\ell^2_{-\mathrm{E}})\,\big[ K_{\mathrm{R}}(-\ell^2_{\mathrm{E}},iM\ell_{\mathrm{E}}z) \nonumber\\
		& - \eta_-Q_{\mathrm{R}}(iQ\ell_{\mathrm{E}}z, MQ) \big] + \sigma^-_{\mathrm{sE}}(\ell^2_{-\mathrm{E}})\,I_{\mathrm{R}},
	\end{align}
\end{subequations}
with $\ell^2_{\pm\mathrm{E}} = {\ell^2_{\mathrm{E}} \mp 2i\eta_\pm Q\ell_{\mathrm{E}}z - \eta^2_\pm Q^2}$. After defining ${\mathbb{U}_{\mathrm{E}}(\ell_{\mathrm{E}}, z, Q)} = {\mathbb{U}(-\ell^2_{\mathrm{E}}, iM\ell_{\mathrm{E}}z, iQ\ell_{\mathrm{E}}z)}$ as the trace matrix in the rest frame, the normalization condition of the BSA then becomes 
\begin{align}
	& 1 = \dfrac{6}{\pi^3} \lim\limits_{Q^2\rightarrow P^2}\dfrac{\partial}{\partial\,Q^2} \int_{0}^{+\infty}d\ell_{\mathrm{E}}\ell^3_{\mathrm{E}} \int_{-1}^{1}dz\sqrt{1-z^2}\, \overline{\mathbb{G}}_{\mathrm{E}}(\ell^2_{\mathrm{E}},z)\nonumber\\
	& \times \mathbb{U}_{\mathrm{E}}(\ell_{\mathrm{E}},z,Q) \mathbb{S}^+_{\mathrm{E}}(\ell_{\mathrm{E}},z,Q) \mathbb{S}^-_{\mathrm{E}}(\ell_{\mathrm{E}},z,Q) \mathbb{G}_{\mathrm{E}}(\ell^2_{\mathrm{E}},z), \label{eq:BSA_normalization_rest_frame}
\end{align}
with the derivative approximated by finite differences and
\begin{equation}
	\overline{\mathbb{G}}_{\mathrm{E}}(\ell^2_{\mathrm{E}},z) = \overline{\mathbb{G}}(-\ell^2_{\mathrm{E}}, iM\ell_{\mathrm{E}}z). \label{eq:def_bbm_G_bar_Euc}
\end{equation}

While in the frame where the bound state carries a finite spacial momentum, we have $P_{\mathrm{E}}^j$ given by Eq.~\eqref{eq:def_bound_state_momentum_boosted_frame} and 
$Q_{\mathrm{E}}^j = ( 0,\, 0,\, Q\,\sinh\varphi,\, -iQ\,\cosh\varphi )$. We then obtain 
\[ \ell_{\mathrm{E}}\cdot Q_{\mathrm{E}} = \ell_{\mathrm{E}} Q\left( y\sqrt{1-z^2}\sinh\varphi - iz\cosh\varphi \right), \] ${Q_{\mathrm{E}}^2 = - Q^2}$, and ${P_{\mathrm{E}}\cdot Q_{\mathrm{E}} = - MQ }$. The matrices for quark propagators become
\begin{equation*}
	\mathbb{S}^\pm_{\mathrm{E}}(\ell_{\mathrm{E}}, y, z, Q) = - \mathbb{S}^\pm(-\ell^2_{\mathrm{E}},-\ell_{\mathrm{E}}\cdot P_{\mathrm{E}}, -\ell_{\mathrm{E}}\cdot Q_{\mathrm{E}}, QM, Q^2),
\end{equation*}
which results in
\begin{subequations}\label{eq:def_Euc_S_pm_boosted_frame}
	\begin{align}
		& \mathbb{S}^+_{\mathrm{E}}(\ell_{\mathrm{E}}, y, z, Q) \!= \!-\sigma^{+}_{\mathrm{vE}}(\ell^2_{+\mathrm{E}}) [ K_{\mathrm{L}}(-\ell^2_{\mathrm{E}}, -\ell_{\mathrm{E}}\cdot P_{\mathrm{E}}, -\ell_{\mathrm{E}}\cdot Q_{\mathrm{E}}) \nonumber\\
		& +\eta_+Q_{\mathrm{L}}(-\ell_{\mathrm{E}}\cdot Q_{\mathrm{E}}, MQ, Q^2) ] + \sigma^{+}_{\mathrm{sE}}(\ell^2_{+\mathrm{E}})\,I_{\mathrm{L}} , \\
		& \mathbb{S}^-_{\mathrm{E}}(\ell_{\mathrm{E}}, y, z, Q) = \sigma^{-}_{\mathrm{vE}}(\ell^2_{-\mathrm{E}}) \,\big[ K_{\mathrm{R}}(-\ell^2_{\mathrm{E}},-\ell_{\mathrm{E}}\cdot P_{\mathrm{E}}) \nonumber\\
		& -\eta_-Q_{\mathrm{R}}(-\ell_{\mathrm{E}}\cdot Q_{\mathrm{E}}, MQ ) \big] + \sigma^{-}_{\mathrm{sE}}(\ell^2_{+\mathrm{E}})\,I_{\mathrm{R}} .
	\end{align}
\end{subequations}
While the trace matrix in the boost frame is defined as
\begin{equation}
	\mathbb{U}_{\mathrm{E}}(\ell_{\mathrm{E}}, y, z, Q) = \mathbb{U}(-\ell^2_{\mathrm{E}}, -\ell_{\mathrm{E}}\cdot P_{\mathrm{E}}, -\ell_{\mathrm{E}}\cdot Q_{\mathrm{E}}). \label{eq:def_bbm_UE_boosted_frame}
\end{equation}
Following these definitions, the normalization condition in terms of the scalar functions of the BSA in the boost frame becomes 
\begin{align}
	& 1 = \dfrac{3}{\pi^3} \lim\limits_{Q^2\rightarrow P^2}\dfrac{\partial}{\partial\,Q^2}\int_{0}^{+\infty} d\ell_{\mathrm{E}}\,\ell^3_{\mathrm{E}}\, \int_{-1}^{1}dz\,\sqrt{1-z^2}\,\int_{-1}^{1}dy \nonumber\\
	& \quad\times \overline{\mathbb{G}}_{\mathrm{E}}(\ell^2_{\mathrm{E}},y,z) \,\mathbb{U}_{\mathrm{E}}(\ell_{\mathrm{E}}, y, z, Q)\, \mathbb{S}^+_{\mathrm{E}}(\ell_{\mathrm{E}}, y, z, Q)\nonumber\\
	& \quad\times \mathbb{S}^-_{\mathrm{E}}(\ell_{\mathrm{E}}, y, z, Q)\,
	\mathbb{G}_{\mathrm{E}}(\ell^2_{\mathrm{E}},y,z), \label{eq:BSA_normalization_boosted_frame}
\end{align}
with $\overline{\mathbb{G}}_{\mathrm{E}}(\ell^2_{\mathrm{E}},y,z) = \overline{\mathbb{G}}(-\ell^2_{\mathrm{E}}, -\ell_{\mathrm{E}} \cdot P_{\mathrm{E}} )$. Similar to Eq.~\eqref{eq:BSA_normalization_rest_frame}, the derivative with respect to $Q^2$ can be approximated numerically by finite differences. 
\subsection{Pion decay constant\label{ss:ps_decay_constant}}
After obtaining the BSA of the pion with proper normalization, the decay constant $f_\pi$ can be calculated from
\begin{equation}
	f_\pi = -\frac{iZ_2 N_{\mathrm{c}}}{m_\pi^2}\int \ddu\ell\ 
	\mathrm{Tr}\!\lf[\gamma_5\slashed{P}\, S(\ell_+)\,\Gamma(\ell,P)\,S(\ell_-)\rg],
	\label{eq:def_fpi}
\end{equation}
with $Z_2$ being the wave function renormalization of the quark propagator and ${\ell_\pm = \ell \pm \eta_\pm P}$~\cite{Tandy:1997qf,Williams:2009wx}. To convert Eq.~\eqref{eq:def_fpi} in terms of the scalar components, let us first apply Eq.~\eqref{eq:ps_BSA_matrix} to the BSA. The multiplications of quark propagators then map the Dirac bases according to Eq.~\eqref{eq:apply_matrix_bbM}. After Dirac traces, Eq.~\eqref{eq:def_fpi} then becomes 
\begin{align}
	f_\pi & = -\dfrac{12i Z_2}{m_\pi^2} \int \ddu\ell 
	\left(0,P^2,\ell\cdot P,0 \right)
	\mathbb{M}(\ell^2, \ell\cdot P)
	\mathbb{G}(\ell^2, \ell\cdot P),
	\label{eq:fpi_reduced}
\end{align}
with $\mathbb{M}(\ell^2,\ell\cdot P)$ givn by Eq.~\eqref{eq:def_matrix_bbM}. 
In the rest frame of the bound state, we obtain
\begin{align}
	f_\pi & = \dfrac{3Z_2}{\pi^3} \int_{0}^{\infty}\dd \ell_{\mathrm{E}}\, \ell^3_{\mathrm{E}}
	\int_{-1}^{+1}\dd z \sqrt{1-z^2} \left(0,1,i\ell_{\mathrm{E}}z/m_\pi,0\right)\ \nonumber\\
	& \quad \times 
	\mathbb{M}^+_{\mathrm{E}}(\ell^2_{\mathrm{E}},z) \, 
	\mathbb{M}^-_{\mathrm{E}}(\ell^2_{\mathrm{E}},z)\, 
	\mathbb{G}_{\mathrm{E}}(\ell^2_{\mathrm{E}},z),
	\label{eq:fpi_Euc_rest_frame}
\end{align}
with ${\mathbb{M}^\pm_{\mathrm{E}}(\ell^2_{\mathrm{E}},z)}$ defined in Eq.~\eqref{eq:def_bbm_pm_euc} and ${\mathbb{G}_{\mathrm{E}}(\ell^2_{\mathrm{E}},z)}$ given by Eq.~\eqref{eq:def_bbm_GE_bsa}. While in the boost frame, Eq.~\eqref{eq:fpi_reduced} becomes
\begin{align}
	& f_\pi = \dfrac{3Z_2}{2\pi^3}\int_{0}^{\infty}\dd \ell_{\mathrm{E}}\,\ell^3_{\mathrm{E}} \int_{-1}^{1}\dd z\,\sqrt{1-z^2}\int_{-1}^{1}\dd y \nonumber\\
	& \! \times \! \left(0,1,-\dfrac{\ell_{\mathrm{E}}\cdot P_{\mathrm{E}}}{m_\pi^2},0\right)
	\mathbb{M}^+_{\mathrm{E}}(\ell^2_{\mathrm{E}},y,z)
	\mathbb{M}^-_{\mathrm{E}}(\ell^2_{\mathrm{E}},y,z) 
	\mathbb{G}_{\mathrm{E}}(\ell^2_{\mathrm{E}}, y, z),
	\label{eq:fpi_Euc_boosted_frame}
\end{align}
with ${\mathbb{M}^\pm_{\mathrm{E}}(\ell^2_{\mathrm{E}},y,z)}$ defined in Eq.~\eqref{eq:def_bbm_pm_euc_boosted_frame} and ${\mathbb{G}_{\mathrm{E}}(k^2_{\mathrm{E}}, y, z)}$ given by Eq.~\eqref{eq:def_bbm_G_E_boosted_frame}. 
Equations~\eqref{eq:fpi_Euc_rest_frame} and \eqref{eq:fpi_Euc_boosted_frame} are suitable for computing the decay constant of the bound state when the BSA is known numerically. 
\subsection{Generalized Gell-Mann--Oakes--Renner relation}
The generalized Gell-Mann--Oakes--Renner relation for the pion of mass $m_{\pi}$ takes the form of
\begin{equation}
	f_\pi\,m_\pi^2 = r_\pi [m_{\mathrm{u}}(\mu^2) + m_{\mathrm{d}}(\mu^2)], 
	\label{eq:def_generalized_GMOR}
\end{equation}
where $m_{\mathrm{u}}$ and $m_{\mathrm{d}}$ are the masses of the up and the down quarks at the renormalization scalar $\mu^2$~\cite{Williams:2009wx}. Here the residue of the pseudoscalar vertex as $r_\pi$ is defined as
\begin{equation}
	r_\pi =  -iZ_2Z_m N_{\mathrm{c}}\,\mathrm{Tr}\int \ddu\ell\ \gamma_5\,S(\ell_+)\Gamma(\ell,P)S(\ell_-), 
	\label{eq:def_rpi}
\end{equation}
with $Z_{\mathrm{m}}$ being the renormalization constant for the quark mass. Similar to the steps taken in deriving Eq.~\eqref{eq:fpi_reduced}, let us first rewrite Eq.~\eqref{eq:def_rpi} in terms of scalar functions of the BSA such that
\begin{equation}
	r_\pi = 12i Z_2 Z_m \int \ddu \ell (1,0,0,0)\, \mathbb{M}(\ell^2,\ell\cdot P)
	\mathbb{G}(\ell^2,\ell\cdot P), 
	\label{eq:rpi_reduced}
\end{equation}
with $\mathbb{M}(\ell^2,\ell\cdot P)$ givn by Eq.~\eqref{eq:def_matrix_bbM}. 

Specifically in the rest frame of the bound state, the expression for vertex residue becomes
\begin{align}
	r_\pi & = -\frac{3\,Z_2 Z_m}{\pi^3}\int_{0}^{\infty}\dd \ell_{\mathrm{E}}\,\ell^3_{\mathrm{E}}\int_{-1}^{1}\dd z\,\sqrt{1-z^2} \nonumber\\
	&\quad \times (1,0,0,0)\,
	\mathbb{M}^+_{\mathrm{E}}(\ell^2_{\mathrm{E}},z)\, 
	\mathbb{M}^-_{\mathrm{E}}(\ell^2_{\mathrm{E}},z)\,
	\mathbb{G}_{\mathrm{E}}(\ell^2_{\mathrm{E}},z),
	\label{eq:rpi_Euc_rest_frame}
\end{align}
with matrices $\mathbb{M}^\pm_{\mathrm{E}}(\ell^2_{\mathrm{E}},z)$ defined in Eq.~\eqref{eq:def_bbm_pm_euc}. While in the boost frame, the vertex residue is given by
\begin{align}
	& r_\pi = -\dfrac{3Z_2 Z_m}{2\pi^3}\int_{0}^{\infty} \dd \ell_{\mathrm{E}}\, \ell^3_{\mathrm{E}}\int_{-1}^{1}\dd z\,\sqrt{1-z^2}\int_{-1}^{1}\dd y \nonumber\\
	& \times (1,0,0,0)\, \mathbb{M}^+_{\mathrm{E}}(\ell^2_{\mathrm{E}}, y, z)\, \mathbb{M}^-_{\mathrm{E}}(\ell^2_{\mathrm{E}}, y, z)\, \mathbb{G}_{\mathrm{E}}(\ell^2_{\mathrm{E}},y,z),
	\label{eq:rpi_Euc_boosted_frame}
\end{align}
with $\mathbb{M}^\pm_{\mathrm{E}}(\ell^2_{\mathrm{E}}, y, z)$ defined by Eq.~\eqref{eq:def_bbm_pm_euc_boosted_frame}. 
\section{PION ELECTROMAGNETIC FORM FACTOR\label{sc:emff}}
\subsection{Ansatz for quark-photon vertex}
\begin{figure*}
	\centering\includegraphics[width=\linewidth]{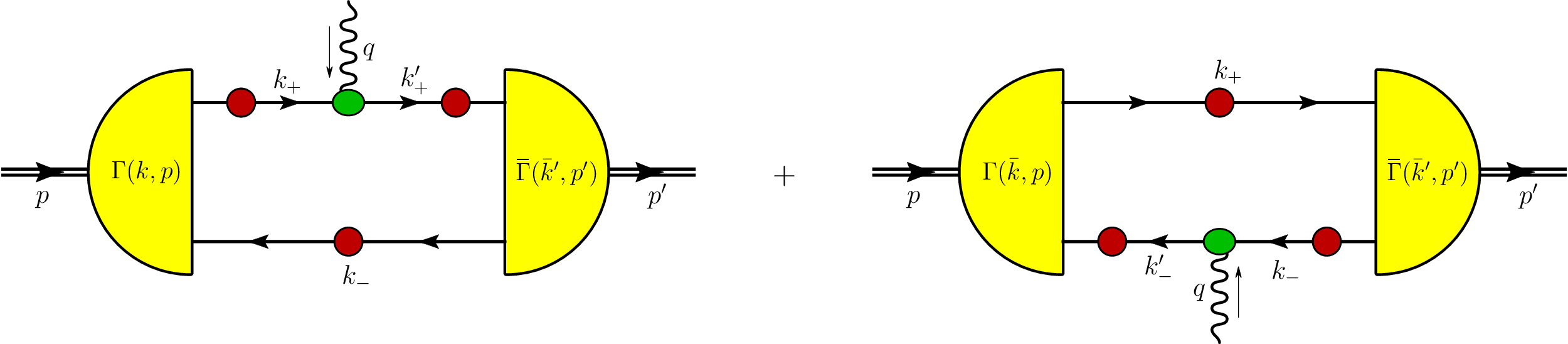}
	\caption{Diagrammatic representations of Eq.~\eqref{eq:em_form_factor_plus_redef} and Eq.~\eqref{eq:em_form_factor_minus_mod} that give the contributions to the EM form factor in the impulse approximation from the valence quark and the valence anti-quark. The yellow blobs on the left and on the right are the BSAs of the bound state. The green blobs with photon lines attached are the quark-photon vertex. The red blobs are the dressed quark propagators.}
	\label{fig:emformfactor}
\end{figure*}
One input to the EM form factor is the quark-photon vertex. Without solving its inhomogeneous BSE, let us resort to the Ansatz from Ref.~\cite{Maris:1999bh} that consists of the Ball-Chiu vertex~\cite{Ball:1980ay} and a transverse piece:
\begin{equation}
	\Gamma^\mu _{\mathrm{EM}}(k,p) = \Gamma^\mu_{\mathrm{BC}}(k,p) + \Gamma^{\mu}_{\mathrm{T}}(k,p), \label{eq:def_EM_vertex}
\end{equation}
with $k$ being the momentum of the quark flowing into the vertex and $p$ that of the outgoing quark. The Ball-Chiu vertex ${\Gamma^\mu_{\mathrm{BC}}(k,p)}$ alone is known to produce a small charge radius~\cite{Maris:1999bh}, and reads
\begin{align}
	\Gamma^{\mu}_{\mathrm{BC}}(k,p) & = 
	\frac{A(k^2)+A(p^2)}{2}\,\gamma^\mu + \frac{A(k^2)-A(p^2)}{k^2 - p^2}\dfrac{t^\mu \sh{t}}{2} \nonumber\\
	& \quad
	+ \dfrac{B(k^2)-B(p^2)}{k^2 - p^2}t^\mu , 
\end{align}
with $t^\mu = k^\mu + p^\mu$. The scalar functions $A(k^2)$ and $B(k^2)$ are defined through Eq.~\eqref{eq:dec_SF}. The transverse part is given by
\begin{equation}
	\Gamma^\mu_{\mathrm{T}}(q-Q/2,q+Q/2) 
	= \left( \gamma^\mu - Q^\mu \sh{Q} / Q^2 \right) C_{\mathrm{MT}}(q^2,Q^2), 
\end{equation}
where we have defined the function ${C_{\mathrm{MT}}(q^2, Q^2)}$ as
\begin{equation}
	C_{\mathrm{MT}}(q^2, Q^2) = \frac{N_{\rho}}{1 + q^4/\omega^4 } \, 
	\frac{f_{\rho} Q^2}{ m_{\rho} (m^2_{\rho} - Q^2) }\,e^{-\alpha(m^2_{\rho} - Q^2)}. 
	\label{eq:def_C_MT}
\end{equation}
In the Euclidean space, this function becomes
\begin{align}
	C_{\mathrm{MT},\mathrm{E}}(q^2_{\mathrm{E}}, Q^2_{\mathrm{E}})
	= \frac{-N_{\rho}}{1 + q^4_{\mathrm{E}}/\omega^4} \, 
	\frac{ f_{\rho} Q^2_{\mathrm{E}} } { m_{\rho} ( m^2_{\rho} + Q^2_{\mathrm{E}} ) }\, 
	e^{-\alpha(m^2_{\rho} + Q^2_{\mathrm{E}})}, 
	\label{eq:def_C_MT_E}
\end{align}
whose parameters are specified in Table~\ref{tab:pmts_MTA_vertex}. The constant $N_\rho$ reflects the normalization of an Ansatz BSA for the $\rho$ meson, which can be determined through the decay constant applying Eq.~\eqref{eq:rho_decay_constant_Euc}~\cite{Maris:1999bh}. 
\begin{table}[b]
	\begin{center}
		\begin{tabular}{ccccc}
			\hline
			$N_\rho$ & $\omega$ & $f_\rho$ & $m_\rho$ & $\alpha$ \\
			\hline
			6.0405 & $0.66~\mathrm{GeV}$ & $201~\mathrm{MeV}$ & $0.875~\mathrm{GeV}$ & $0.1~(\mathrm{GeV})^{-2}$ \\
			\hline
		\end{tabular}
	\end{center}
	\caption{Parameters for the transverse Ansatz of the quark-photon vertex in Eq.~\eqref{eq:def_C_MT_E}. The normalization $N_{\rho}$ is determined from the decay constant of the $\rho$ meson $f_{\rho}$ as in Appendix~\ref{sc:n_rho}. } \label{tab:pmts_MTA_vertex}
\end{table}
\subsection{Form Factor Contribution from Valence Quark}
Denote the valence quark contribution to the EM form factor of pseudoscalar bound states by $F_+(q^2)$. When the photon momentum is $q^\mu$, in the impulse approximation we have
\begin{align}
	& P^\mu F_{+}(q^2) = -ie_q\,\mathrm{Tr}_{\mathrm{cD}}\,\int d\underline{l}\, 
	\overline{\Gamma}\left(\kappa',\, - \Pi'\right)\,
	S^{+}(\kappa'_+) \nonumber\\
	& \times
	\Gamma_{\mathrm{+EM}}^\mu\left(\kappa_+, \, \kappa'_+ \right)\, 
	S^{+}(\kappa_+)\,
	\Gamma\left(\kappa, \, \Pi \right)\,
	S^{-}(\kappa_-),\label{eq:em_form_factor_plus_redef}
\end{align}
where $e_q$ is the charge of the valence quark in units of the elementary charge~\cite{Maris:1999bh,Maris:2000sk}. Equation~\eqref{eq:em_form_factor_plus_redef} is diagrammatically represented in Fig.~\ref{fig:emformfactor} with traces operating in the color and Dirac spaces. Functions $S^+(p)$ and $S^-(p)$ are respectively the dressed propagators of the valence quark and antiquark. Let us then define momenta $\kappa = l - \eta_- q/2$, $\Pi = P - q / 2$, $\kappa' = l + \eta_- q / 2$, and $\Pi' = P + q / 2$ so that quark momenta in Eq.~\eqref{eq:em_form_factor_plus_redef} are then given by
\begin{subequations}\label{eq:def_kappa_pm_kappap_pm}
	\begin{align}
		\kappa_\pm & = \kappa \pm \eta_{\pm}\Pi, \\
		\kappa'_\pm & = \kappa'\pm \eta_{\pm}\Pi' .
	\end{align}
\end{subequations}
with $\eta_+ + \eta_- = 1$. Because the bound states are on-shell, we also have ${( P \pm q/2 ) ^2} = {M^2}$ with $P\cdot q = 0$. 
The conjugate BSA is defined in Eq.~\eqref{eq:Gamma_bar_neg_P}. Additionally, $\Gamma^\mu_{+\mathrm{EM}}(\kappa_+, \kappa'_+)$ is the quark-photon vertex. 

With the assistance of Eq.~\eqref{eq:Gamma_bar_neg_P}. the multiplication of quark propagator to the BSA in the first line of Eq.~\eqref{eq:em_form_factor_plus_redef} becomes
\begin{align}
	& \overline{\Gamma}(\kappa',-\Pi') \, S^+(\kappa'_+)
	\nonumber\\
	& = \overline{\mathbb{G}}(\kappa'^2,\kappa'\cdot \Pi')\, \mathbb{N}^+(\kappa'^2,\kappa'\cdot \Pi') 
	\overline{T}(\kappa',\Pi')
	\gamma_5, \label{eq:S_Fp_Gamma_bar}
\end{align}
where we have defined the matrix $\mathbb{N}^+(\kappa'^2, \kappa'\cdot \Pi')$ as
\begin{align}
	& \mathbb{N}^+(\kappa'^2, \kappa'\cdot \Pi') = \big[ - k_{\mathrm{R}}^{\mathrm{T}}(\kappa'^2, \kappa'\cdot \Pi') \nonumber\\
	& - \eta_+ P_{\mathrm{R}}^{\mathrm{T}}(\kappa'\cdot \Pi',\Pi'^2) \big] \sigma^+_{\mathrm{v}}(\kappa'^2_+) + \mathbb{1}_4 \sigma^+_{\mathrm{s}}(\kappa'^2_+).  \label{eq:def_Np_bbm} 
\end{align}
Transposes are taken in Eq.~\eqref{eq:def_Np_bbm} for matrices defined in Eq.~\eqref{eq:multi_rule_LR} because the Dirac bases in Eq.~\eqref{eq:S_Fp_Gamma_bar} form a column vector. Meanwhile, the application of Eq.~\eqref{eq:ps_BSA_matrix} with Eq.~\eqref{eq:apply_matrix_bbM} gives the scalar decomposition of the BS wave function. Equation~\eqref{eq:em_form_factor_plus_redef} then becomes
\begin{align}
	& P^\mu F_{+}(q^2)
	= ie_qN_\mathrm{c}\,\int d\underline{l}\,
	\overline{\mathbb{G}}(\kappa'^2,\kappa'\cdot \Pi')\,\mathbb{N}^{+}(\kappa'^2,\kappa'\cdot \Pi') \nonumber\\
	& \times \mathrm{Tr}_{\mathrm{D}}\, \overline{T}(\kappa',\Pi')
	\gamma_5\Gamma_{\mathrm{+EM}}^\mu\left(\kappa_+, \, \kappa'_+ \right)\!\gamma_5
	T(\kappa,\Pi) \nonumber\\
	& \times \mathbb{M}(\kappa^2,\kappa\cdot \Pi)\, \mathbb{G}(\kappa^2,\kappa\cdot \Pi), 
	\label{eq:EMFF_quark_intermediate}
\end{align}
with the matrix $\mathbb{M}(\kappa^2,\kappa\cdot \Pi)$ defined in Eq.~\eqref{eq:def_matrix_bbM}. The trace operations in Eq.~\eqref{eq:EMFF_quark_intermediate} depend on the structure of the EM vertex. Specifically with the Ansatz vertex in Eq.~\eqref{eq:def_EM_vertex}, we need to calculate 
\begin{equation}
	\mathbb{T}_{0}(P, q, l) = \dfrac{1}{4}\,\mathrm{Tr}\,\overline{T}(\kappa',\Pi')\,T(\kappa,\Pi) \label{eq:def_mbb_T0}
\end{equation}
and
\begin{equation}
	\mathbb{T}_{1-}^\mu(P, q, l) = \dfrac{1}{4}\,\mathrm{Tr}\,\overline{T}(\kappa',\Pi')\,\gamma^\mu\,T(\kappa,\Pi)
	\label{eq:def_mbb_T1m}
\end{equation}
with $T(k,P)$ and $\overline{T}(k,P)$ given by Eq.~\eqref{eq:def_Tj_basis} and Eq.~\eqref{eq:def_Tj_basis_transpose}. Substituting these trace matrices into Eq.~\eqref{eq:EMFF_quark_intermediate} produces the following expression of $F_+(q^2)$ in terms of the scalar components of the BSA:
\begin{align}
	& F_{+}(q^2) = \dfrac{4ie_q}{P^2}N_\mathrm{c}\,\int d\underline{l}\, \overline{\mathbb{G}}(\kappa'^2,\kappa'\cdot \Pi')\, \mathbb{N}^{+}(\kappa'^2,\kappa'\cdot \Pi') \nonumber\\
	& \times\bigg\{ -\left[ P \cdot \mathbb{T}_{1-}(P, q, l) \right] \bigg[ \dfrac{A^+(\kappa^2_+) + A^+(\kappa'^2_+)}{2} \nonumber\\
	& \quad + C_{\mathrm{MT}}\left(\left(l+\eta_+P\right)^2,q^2\right) \bigg] - 2\left[(l+\eta_+P)\cdot P\right] \nonumber\\
	& \quad \times \,\left[ (l+\eta_+P)\cdot \mathbb{T}_{1-}(P, q, l)\right] \left[ \dfrac{A^+(\kappa^2_+) - A^+(\kappa'^2_+)}{\kappa^2_+ - \kappa'^2_+} \right] \nonumber\\
	& \quad + 2\left[(l+\eta_+P)\cdot P \right] \mathbb{T}_{0}(P, q, l)\, \dfrac{B^+(\kappa^2_+) - B^+(\kappa'^2_+)}{\kappa^2_+ - \kappa'^2_+} \bigg\} \nonumber\\
	& \times \mathbb{M}(\kappa^2,\kappa\cdot \Pi)\, \mathbb{G}(\kappa^2,\kappa\cdot \Pi) , \label{eq:emff_quark_Minkowski}
\end{align}
where the contraction with $P_{\mu}$ has been applied. Here $\kappa_+$ and $\kappa'_+$ are specified in Eq.~\eqref{eq:def_kappa_pm_kappap_pm}. We have applied $P\cdot q = 0$ due to the bound state being on-shell. For any momentum $k^\mu$, we also have ${k\cdot \mathbb{T}_{1-}(P, q, l) = k_\mu \cdot \mathbb{T}^{\mu}_{1-}(P, q, l)}$.
\subsection{Form Factor Contribution from Valence Anti-quark}
The anti-quark contribution to the EM form factor as $ F_{-}(q^2)$ in the impulse approximation is given by
\begin{align}
	& P^\mu F_{-}(q^2) = -ie_{\bar{q}}\,\mathrm{Tr}_{\mathrm{cD}}\,\int d\underline{l}\, S^-(\kappa_-')\, \overline{\Gamma}(\kappa',-\Pi')\nonumber\\
	& \times S^+(\kappa_+)\, \Gamma(\kappa, \Pi)\, S^-(\kappa_-)\, \Gamma^\mu_{-\mathrm{EM}}(\kappa'_-,\kappa_-) ,\label{eq:em_form_factor_minus_mod}
\end{align}
where $e_{\bar{q}}$ is the charge of the anti-quark in units of the elementary charge~\cite{Maris:1999bh,Maris:2000sk}. While $\Gamma_{\mathrm{-EM}}^\mu(\kappa'_-,\kappa_-)$ is the anti-quark-photon vertex, which is identical to $\Gamma_{\mathrm{+EM}}^\mu(\kappa_+,\kappa'_+)$ in Eq.~\eqref{eq:em_form_factor_plus_redef} for the pion in the isospin symmetric limit. Equation~\eqref{eq:em_form_factor_minus_mod} is diagrammatically represented in Fig.~\ref{fig:emformfactor}. After defining $\kappa = l + \eta_+ q / 2$, $\Pi = P - q / 2$, $\kappa' = l - \eta_+q / 2$, and $\Pi' = P + q / 2$, quark momenta in Eq.~\eqref{eq:em_form_factor_minus_mod} are given by Eq.~\eqref{eq:def_kappa_pm_kappap_pm}. 

Following steps similar to those in the derivation of Eq.~\eqref{eq:emff_quark_Minkowski}, let us decompose the antiquark contribution to the EM form factor into the scalar components of the BSA. When the normalization of the BSA is given by Eq.~\eqref{eq:BSA_normalization} with the EM vertex satisfying the Ward identity, the conservation of electric charge is satisfied by Eq.~\eqref{eq:em_form_factor_plus_redef} and Eq.~\eqref{eq:em_form_factor_minus_mod} at $q^2=0$~\cite{Maris:2000sk}. The decomposition of ${S^-(\kappa'_-)}\,{\overline{\Gamma}(\kappa',-\Pi')}$ is
\begin{align}
	& S^-(\kappa'_-) \, \overline{\Gamma}(\kappa',-\Pi')\nonumber\\ 
	& = \overline{\mathbb{G}}(\kappa'^2, \kappa'\cdot \Pi') \,
	\mathbb{N}^-(\kappa'^2, \kappa'\cdot \Pi')
	\overline{T}(\kappa',\Pi')
	\,\gamma_5 ,
\end{align}
after defining $\mathbb{N}^-(\kappa'^2, \kappa'\cdot \Pi')$ according to
\begin{align}
	\mathbb{N}^-(\kappa'^2, \kappa'\cdot \Pi') & = \big[ k_{\mathrm{L}}^{\mathrm{T}}(\kappa'^2, \kappa'\cdot \Pi')-\eta_-P_{\mathrm{L}}^{\mathrm{T}}(\kappa'\cdot \Pi', \Pi'^2) \big] \nonumber\\
	& \quad\times \sigma^-_{\mathrm{v}}(\kappa'^2_-) + \mathbb{1}_{4}\sigma^-_{\mathrm{s}}(\kappa'^2_-)  \label{eq:def_bbm_n_antiquark}
\end{align} 
with $P_{\mathrm{L}}^{\mathrm{T}}$ and $k_{\mathrm{L}}^{\mathrm{T}}$ given by Eq.~\eqref{eq:multi_rule_LR}. Further based on Eq.~\eqref{eq:ps_BSA_matrix}
and Eq.~\eqref{eq:apply_matrix_bbM}, Eq.~\eqref{eq:em_form_factor_minus_mod} becomes 
\begin{align}
	& P^\mu F_{-}(q^2) = ie_{\bar{q}}\,N_{\mathrm{c}}\,\int d\underline{l} \, 
	\overline{\mathbb{G}}(\kappa'^2,\kappa'\cdot \Pi')\,\mathbb{N}^-(\kappa'^2,\kappa'\cdot \Pi') \nonumber\\
	& \times \mathrm{Tr}_{\mathrm{D}}
	\overline{T}(\kappa',\Pi')
	T(\kappa,\Pi)
	\Gamma^\mu_{-\mathrm{EM}}(\kappa'_-,\kappa_-) 
	\nonumber\\
	& \times \mathbb{M}(\kappa^2,\kappa\cdot \Pi) \, \mathbb{G}(\kappa^2,\kappa\cdot \Pi).
\end{align}
When at most one $\gamma$-matrix is present in the EM vertex of the antiquark, the relevant trace operations are represented by Eq.~\eqref{eq:def_mbb_T0} and 
\begin{equation}
	\mathbb{T}_{1+}^\mu(P, q, l) = \dfrac{1}{4}\,\mathrm{Tr}\,\overline{T}(\kappa',\Pi')\,T(\kappa,\Pi) \,\gamma^\mu. \label{eq:def_mbb_T1p}
\end{equation}
Applying the Ansatz quark-photon vertex in Eq.~\eqref{eq:def_EM_vertex}, $F_-(q^2)$ in terms of the BSA becomes
\begin{align}
	& F_-(q^2) = \dfrac{4ie_{\bar{q}}}{P^2}\,N_{\mathrm{c}}\,\int d\underline{l}\, 
	\overline{\mathbb{G}}(\kappa'^2,\kappa'\cdot \Pi') \, \mathbb{N}^-(\kappa'^2,\kappa'\cdot \Pi') \nonumber\\
	& \times \bigg\{
	[P\cdot \mathbb{T}_{1+}(P, q, l)]\,\bigg[ \dfrac{A^-(\kappa'^2_-) + A^-(\kappa^2_-)}{2} \nonumber\\
	& \quad + C_{\mathrm{MT}}\left( (l-\eta_-P)^2,q^2 \right) \bigg] + 2[(l-\eta_-P)\cdot P] \nonumber\\
	& \quad \times [(l-\eta_-P)\cdot \mathbb{T}_{1+}(P, q, l)]  \left[\dfrac{A^-(\kappa'^2_-) - A^-(\kappa^2_-)}{\kappa'^2_- -\kappa^2_-} \right] \nonumber\\
	& \quad + 2[(l-\eta_-P)\cdot P]\, \mathbb{T}_{0}(P, q, l)\dfrac{B^-(\kappa'^2_-) - B^-(\kappa^2_-)}{\kappa'^2_- - \kappa^2_-}
	\bigg\} \nonumber\\
	& \times \mathbb{M}(\kappa^2,\kappa\cdot \Pi) \,
	\mathbb{G}(\kappa^2,\kappa\cdot \Pi), \label{eq:emff_antiquark_Minkowski}
\end{align}
with ${k\cdot \mathbb{T}_{1+}(P, q, l) = k_\mu \cdot \mathbb{T}^{\mu}_{1+}(P, q, l)}$ for any $k^\mu$. In deriving Eq.~\eqref{eq:emff_antiquark_Minkowski}, we have also applied Eq.~\eqref{eq:def_kappa_pm_kappap_pm} and ${P\cdot q} = 0$. 
\subsection{Euclidean Space Kinematics in Center-of-Mass Frame}
The expressions for the EM form factor as Eq.~\eqref{eq:emff_quark_Minkowski} and Eq.~\eqref{eq:emff_antiquark_Minkowski} await to be converted into the Euclidean space before computation. Because the bound states with momenta ${P\pm q/2}$ are on-shell, we have $P^2 = M^2 + {Q^2/4}$ and $P\cdot q = 0$. Here we have introduced $Q^2 = q_{\mathrm{E}}^2 = {-q^2}$ as the square of the spacelike photon momentum. This definition is different from the timelike $Q^2$ in the normalization condition in Sec.~\ref{sc:static}. Let us then choose a reference frame where the photon momentum is parallel to the $z$-axis such that
\begin{subequations}
	\label{eq:def_transverse_com}
	\begin{align}
		q_{\mathrm{E}}^j & = \left(0, 0, Q, 0 \right), \\
		P_{\mathrm{E}}^j & = \left(0,\,0,\,0,\,-i\sqrt{M^2 + Q^2/4}\right). 
	\end{align}
\end{subequations}
Because the transfer of the spacial momentum from the photon only occurs in the longitudinal direction, this reference frame is named ``transverse center-of-mass frame''. The loop momenta in Eq.~\eqref{eq:emff_quark_Minkowski} and Eq.~\eqref{eq:emff_antiquark_Minkowski} are parameterized in analogue with Eq.~\eqref{eq:def_loop_momentum}, resulting in the integration measure similar to Eq.~\eqref{eq:itg_measure_dkp_ori}. 

Within this frame the quark propagators are sampled in the complex-momentum plane by Eq.~\eqref{eq:emff_quark_Minkowski} and Eq.~\eqref{eq:emff_antiquark_Minkowski}. In the case of $F_+(q^2)$, the momenta of quarks interacting with the photon are given by $\kappa_+$ and $\kappa'_+$ in  Eq.~\eqref{eq:def_kappa_pm_kappap_pm}, which correspond to
\begin{align}
	& \left( l + \eta_+ P \pm q/2 \right)^2_{\mathrm{E}}  = l^2_{\mathrm{E}} - 2i\eta_+\sqrt{M^2 + Q^2/4}\,l_{\mathrm{E}}z_l \nonumber\\
	& \quad \pm Ql_{\mathrm{E}}y_l\sqrt{1-z^2_l} - \eta_+^2\left(M^2 + Q^2/4 \right) +Q^2/4
\end{align}
in the Euclidean space. While the momentum of the spectator quark as $\kappa_-=\kappa'_-$ in Eq.~\eqref{eq:def_kappa_pm_kappap_pm} becomes
\begin{align}
	(l-\eta_- P)^2_{\mathrm{E}} & = l_{\mathrm{E}}^2 + 2i\eta_-\sqrt{M^2 + Q^2/4}\,l_{\mathrm{E}}z_l \nonumber\\ & \quad - \eta_-^2\left( M^2 + Q^4/4 \right) .
\end{align}
These momenta fall within the parabolic regions sampled by the BSE in the boost frame. 

While in the case of $F_-(q^2)$, the momenta of interacting quarks are given by $\kappa'_{-}$ and $\kappa_{-}$, which in the Euclidean space become
\begin{align}
	& (l - \eta_- P \pm q/2)^2_{\mathrm{E}} = l^2_{\mathrm{E}} + 2i \eta_- \sqrt{M^2 + Q^2/4} \, l_{\mathrm{E}}z_l \nonumber\\
	& \pm Ql_{\mathrm{E}}y_l \sqrt{1-z^2_l} - \eta_-^2\left(M^2+Q^2/4\right) + Q^2/4 .
\end{align}
While the momentum of the spectator quark as $\kappa'_{+} = \kappa_+$ is given by
\begin{align}
	(l + \eta_+P)^2_{\mathrm{E}} & = l^2_{\mathrm{E}} - 2i\eta_+\sqrt{M^2+Q^2/4} \, l_{\mathrm{E}} z_l \nonumber\\
	& \quad - \eta^2_+\left( M^2 + Q^2/4 \right) .
\end{align}
Similar to $F_{+}(q^2)$, these momenta also fall within the parabolic regions sampled by the boost-frame BSE. 

\subsection{Euclidean Space  BSAs sampled by the EM form factor}
We would like to explore if both Eq.~\eqref{eq:emff_quark_Minkowski} and Eq.~\eqref{eq:emff_antiquark_Minkowski} in the Euclidean space require scalar functions of the BSA in the boost frames. Based on Eq.~\eqref{eq:def_kappa_pm_kappap_pm}, the momenta of these BSAs are given by
\begin{subequations}
	\label{eq:def_kappa_pi_emff}
	\begin{align}
		\kappa & = l \pm \eta\, q / 2 \\
		\Pi & = P \pm q/2
	\end{align}
\end{subequations}
with $\eta \in \{\eta_+,\, \eta_-\}$, $\kappa \in \{ \kappa, \, \kappa'\}$, and $\Pi \in \{ \Pi, \, \Pi'\}$. Because the bound state is on-shell, we also have ${ \Pi^2_{\mathrm{E}} = - M^2 }$. Within the transverse center-of-mass frame specified by Eq.~\eqref{eq:def_transverse_com}, $\kappa^2$ in the Euclidean space becomes
\begin{equation}
	\kappa_{\mathrm{E}}^2 = l^2_{\mathrm{E}} \pm \eta Ql_{\mathrm{E}}y_l\sqrt{1-z^2_l} + \eta^2\,Q^2/4 . \label{eq:def_kappa_E_squared}
\end{equation}
We also obtain the following inner product
\begin{equation}
	\kappa_{\mathrm{E}}\cdot \Pi_{\mathrm{E}} = -i\sqrt{M^2 + \dfrac{Q^2}{4}}l_{\mathrm{E}}z_l \pm \dfrac{Q}{2}l_{\mathrm{E}} y_l\sqrt{1-z^2_l} + \dfrac{\eta}{4}Q^2. \label{eq:kappa_dot_Pi_Euc}
\end{equation}

The scalar functions of the BSA in the rest frame depend on the radial momentum squared and the temporal angle. The radial momentum squared as the first variable has a unique definition by Eq.~\eqref{eq:def_kappa_E_squared}. Let us then consider the following $2$ extensions of the temporal angle in the rest-frame BSA as an attempt to account for the nonzero spacial momenta of bound states in the EM form factor. 

First let the second variable $\tilde{z}$ be defined as the cosine of an angle by $\tilde{z} = i {\kappa_{\mathrm{E}} \cdot \Pi_{\mathrm{E}}} / {(\kappa_{\mathrm{E}} M)} $. In the limit of $Q\rightarrow 0$ we have $\tilde{z} = z_l$, where the rest-frame BSA contains all the information to compute the EM form factor. However when $Q\neq 0$, $\tilde{z}$ becomes complex with $-[1+Q^2/(4M^2)]^{1/2} \leq \mathrm{Re}\{\tilde{z}\} \leq [1+Q^2/(4M^2)]^{1/2}$. Since the variable $z$ in Eq.~\eqref{eq:ps_BSE_Euc_rest_frame} is only defined within $[-1,1]$, an extrapolation is needed in order to compute the EM form factor. Therefore the rest-frame BSA does not contain all the information required. 

An alternative definition of the variable $\tilde{z}$ is the cosine of the angle between the vector $\kappa^j_{\mathrm{E}}$ and the Euclidean temporal axis by $\tilde{z} = \kappa^4_{\mathrm{E}}/\kappa_{\mathrm{E}}$. In the limit of $Q\rightarrow 0$, we recover $\tilde{z} = z_l$. Furthermore when $Q\neq 0$, $\tilde{z}$ falls within $[-1,\,1]$. However when the spacial momentum of the bound state is nonzero, the scalar functions of the BSA depend on a spacial angle of $\kappa_{\mathrm{E}}^j$ not captured by $\tilde{z}$. The rest-frame BSA is therefore only applicable when such a dependence can be neglected. Because our numeric solutions in the boost frame indicate otherwise, this extension of the temporal angle is an approximation when applied to compute the EM form factor. 

In both scenarios, the BSA in the rest frame does not contain all the information of the bound state to compute the EM form factor with off-shell photons. Alternatively when the BSE at finite rapidity as Eq.~\eqref{eq:ps_BSE_Euc_boosted_frame} is solved, the resulting scalar functions automatically depend on $3$ variables, allowing native support to compute the EM form factor. Specifically for a given loop momentum, Eq.~\eqref{eq:def_kappa_pi_emff} gives
\begin{align}
	\kappa_{\mathrm{E}}^j & = \Big(
	l_{\mathrm{E}}\sqrt{(1-y^2_l)(1-z^2_l)}\cos\phi_l,\, l_{\mathrm{E}}\sqrt{(1-y^2_l)(1-z^2_l)} \nonumber\\ & \quad \times \sin\phi_l,\, l_{\mathrm{E}}y_l\sqrt{1-z^2_l} \pm \eta Q/2,\, l_{\mathrm{E}}z_l \Big) .
\end{align}
While as a variable of the BSA, we have 
\begin{align}
	\kappa_{\mathrm{E}}^j & = \big(	\kappa_{\mathrm{E}}\sqrt{(1-y^2)(1-z^2)}\cos\phi,\, \kappa_{\mathrm{E}}\sqrt{(1-y^2)(1-z^2)} \nonumber\\ 
	& \quad \times \sin\phi,\, \kappa_{\mathrm{E}}y\sqrt{1-z^2},\, \kappa_{\mathrm{E}}z \big). 
\end{align}	
These $2$ identities result in the following mapping:
\begin{subequations}\label{eq:varable_mapping_l_to_kappa}
	\begin{align}
		\kappa_{\mathrm{E}}^2 & = l_{\mathrm{E}}^2 \pm \eta Q l_{\mathrm{E}}y_l\sqrt{1-z^2_l} + \eta^2 Q^2/4, \\
		z & = l_{\mathrm{E}} z_l / \kappa_{\mathrm{E}}, \\
		y & = \left( l_{\mathrm{E}} y_l\sqrt{1-z^2_l} \pm \eta Q/2 \right) / \sqrt{\kappa_{\mathrm{E}}^2 - l_{\mathrm{E}}^2 z_l^2} . 
	\end{align} 
\end{subequations}
Additionally, scalar functions of the pion BSA are independent of the azimuthal angle $\phi$. Consequently the right-hand sides of Eq.~\eqref{eq:emff_quark_Minkowski} and Eq.~\eqref{eq:emff_antiquark_Minkowski} do not depend on this angle.  Equation~\eqref{eq:varable_mapping_l_to_kappa} then specifies how the a given loop momentum in the EM form factor inquires the BSA scalar functions. It also indicates the requirement of interpolation when these scalar functions are only known on a chosen grid of $\kappa_{\mathrm{E}}^2$, $y$, and $z$. Such a dependence can be alternatively solved from the BSE in Eq.~\eqref{eq:ps_BSE_Euc_boosted_frame}. 

\subsection{EM form factor in the Euclidean space}
Let us define the EM form factors in the Euclidean space as
\begin{equation}
	F_{\pm_{\mathrm{E}}}(Q^2) = F_\pm(-q^2_{\mathrm{E}}),\label{eq:def_Euc_emff}
\end{equation}
with $F_\pm(-q^2_{\mathrm{E}})$ given by Eq.~\eqref{eq:emff_quark_Minkowski} and Eq.~\eqref{eq:emff_antiquark_Minkowski}. When the photon momentum is referred as $Q^2$, the function on the left-hand side of Eq.~\eqref{eq:def_Euc_emff} is automatically selected, which is equivalent to $F_{\pm}(Q^2) = {F_{\pm_{\mathrm{E}}}(Q^2)}$. For the computation of the EM form factor, we would like to convert Eq.~\eqref{eq:emff_quark_Minkowski} and Eq.~\eqref{eq:emff_antiquark_Minkowski} in the Euclidean space. 

In order to represent the trace operations in the EM from factor, let us define the Euclidean-space versions of Eq.~\eqref{eq:def_mbb_T0}, Eq.~\eqref{eq:def_mbb_T1p}, and Eq.~\eqref{eq:def_mbb_T1m} as
\begin{align}
	& \mathbb{T}_{0\mathrm{E}}(\Pi'_{\mathrm{E}}\cdot \Pi_{\mathrm{E}}, \Pi'_{\mathrm{E}}\cdot \kappa_{\mathrm{E}}, \kappa'_{\mathrm{E}}\cdot \Pi_{\mathrm{E}}, \kappa'_{\mathrm{E}}\cdot \kappa_{\mathrm{E}}) \nonumber\\
	& = \begin{pmatrix}
		1 & 0 & 0 & 0 \\
		0 & -\Pi'_{\mathrm{E}}\cdot \Pi_{\mathrm{E}} & -\Pi'_{\mathrm{E}}\cdot \kappa_{\mathrm{E}} & 0 \\
		0 & -\kappa'_{\mathrm{E}}\cdot \Pi_{\mathrm{E}} & -\kappa'_{\mathrm{E}}\cdot \kappa_{\mathrm{E}} & 0 \\
		0 & 0 & 0 & t_{44}
	\end{pmatrix} \label{eq:def_bbf_T0E}
\end{align}
and
\begin{align}
	& \mathbb{T}_{1\pm\mathrm{E}}( \Pi'_{\mathrm{E}}\cdot \Pi_{\mathrm{E}}, \Pi'_{\mathrm{E}}\cdot \kappa_{\mathrm{E}}, \kappa'_{\mathrm{E}}\cdot \Pi_{\mathrm{E}}, \kappa'_{\mathrm{E}}\cdot \kappa_{\mathrm{E}}, \nonumber\\
	& \quad\quad\quad \Pi'_{\mathrm{E}}\cdot P_{\mathrm{E}}, \Pi_{\mathrm{E}}\cdot P_{\mathrm{E}}, \kappa'_{\mathrm{E}}\cdot P_{\mathrm{E}}, \kappa_{\mathrm{E}}\cdot P_{\mathrm{E}} ) \nonumber\\
	& = \begin{pmatrix}
		0 & -\Pi_{\mathrm{E}} \cdot P_{\mathrm{E}} & - \kappa_{\mathrm{E}} \cdot P_{\mathrm{E}} & 0 \\
		- \Pi'_{\mathrm{E}}\cdot P_{\mathrm{E}} & 0 & 0 & \pm t_{24} \\
		- \kappa'_{\mathrm{E}}\cdot P_{\mathrm{E}}  & 0 & 0 & \pm t_{34} \\
		0 & \pm t_{42} & \pm t_{43} & 0
	\end{pmatrix} \label{eq:def_bbf_T1pmE}
\end{align}
with
\begin{subequations}
	\begin{align}
		t_{24} & = -\Pi'_{\mathrm{E}}\cdot \Pi_{\mathrm{E}}\, \kappa_{\mathrm{E}}\cdot P_{\mathrm{E}} + \Pi'_{\mathrm{E}}\cdot \kappa_{\mathrm{E}}\, \Pi_{\mathrm{E}}\cdot P_{\mathrm{E}}, \\
		t_{34} & = - \kappa'_{\mathrm{E}}\cdot\Pi_{\mathrm{E}}\,\kappa_{\mathrm{E}}\cdot P_{\mathrm{E}} + \kappa'_{\mathrm{E}}\cdot\kappa_{\mathrm{E}}\, \Pi_{\mathrm{E}}\cdot P_{\mathrm{E}}, \\
		t_{42} & = -\kappa'_{\mathrm{E}}\cdot \Pi_{\mathrm{E}}\, \Pi'_{\mathrm{E}}\cdot P_{\mathrm{E}} + \Pi'_{\mathrm{E}}\cdot \Pi_{\mathrm{E}} \, \kappa'_{\mathrm{E}}\cdot P_{\mathrm{E}}, \\
		t_{43} & = -\kappa_{\mathrm{E}}\cdot \kappa'_{\mathrm{E}}\, \Pi'_{\mathrm{E}}\cdot P_{\mathrm{E}} + \kappa_{\mathrm{E}}\cdot \Pi'_{\mathrm{E}}\, \kappa'_{\mathrm{E}}\cdot P_{\mathrm{E}}, \\
		t_{44} & = \kappa'_{\mathrm{E}} \cdot \Pi_{\mathrm{E}}\, \kappa_{\mathrm{E}}\cdot \Pi'_{\mathrm{E}} - \kappa'_{\mathrm{E}}\cdot \kappa_{\mathrm{E}}\, \Pi'_{\mathrm{E}}\cdot \Pi_{\mathrm{E}} . 
	\end{align}
\end{subequations}
The expression of the EM form factor from the valence quark as Eq.~\eqref{eq:emff_quark_Minkowski} then becomes
\begin{align}
	& F_{+}(Q^2) = -\dfrac{e_q N_{\mathrm{c}}}{2\pi^3 P^2_{\mathrm{E}}} \int_{0}^{+\infty}dl_{\mathrm{E}} \, l^3_{\mathrm{E}}\int_{-1}^{1}dy_{\mathrm{l}}\, \int_{-1}^{1}dz_{\mathrm{l}}\,\sqrt{1-z^2_{\mathrm{l}}} \nonumber\\	
	& \times
	\overline{\mathbb{G}}_{\mathrm{E}}(\kappa'^2_{\mathrm{E}},y',z')
	\mathbb{N}^{+}_{\mathrm{E}}(\kappa'^2_{\mathrm{E}},\kappa'_{\mathrm{E}}\cdot \Pi'_{\mathrm{E}}) \bigg\{ - \mathbb{T}_{1-\mathrm{E}} ( \Pi'_{\mathrm{E}}\cdot \Pi_{\mathrm{E}}, \nonumber\\
	& \Pi'_{\mathrm{E}}\cdot \kappa_{\mathrm{E}}, \kappa'_{\mathrm{E}}\cdot \Pi_{\mathrm{E}}, \kappa'_{\mathrm{E}}\cdot \kappa_{\mathrm{E}}, \Pi'_{\mathrm{E}}\cdot P_{\mathrm{E}}, \Pi_{\mathrm{E}}\cdot P_{\mathrm{E}}, \kappa'_{\mathrm{E}}\cdot P_{\mathrm{E}}, \kappa_{\mathrm{E}}\cdot P_{\mathrm{E}} ) \nonumber\\
	& \times \left[ \dfrac{A^+_{\mathrm{E}}(\kappa^2_{+\mathrm{E}}) + A^{+}_{\mathrm{E}}(\kappa'^2_{+\mathrm{E}}) }{2} + C_{\mathrm{MT},\mathrm{E}}( L^2_{+\mathrm{E}}, Q^2 ) \right] \nonumber\\
	& - 2 (L_{+\mathrm{E}}\cdot P_{\mathrm{E}})\,\mathbb{T}_{1-\mathrm{E}} 
	( \Pi'_{\mathrm{E}}\cdot \Pi_{\mathrm{E}}, \Pi'_{\mathrm{E}}\cdot \kappa_{\mathrm{E}}, \kappa'_{\mathrm{E}}\cdot \Pi_{\mathrm{E}}, \kappa'_{\mathrm{E}}\cdot \kappa_{\mathrm{E}},  \nonumber\\
	& \Pi'_{\mathrm{E}}\cdot L_{+\mathrm{E}}, \Pi_{\mathrm{E}}\cdot L_{+\mathrm{E}}, \kappa'_{\mathrm{E}}\cdot L_{+\mathrm{E}}, \kappa_{\mathrm{E}}\cdot L_{+\mathrm{E}} ) \nonumber \\
	& \times
	\dfrac{A^+_{\mathrm{E}}(\kappa^2_{+\mathrm{E}}) - A^{+}_{\mathrm{E}}(\kappa'^2_{+\mathrm{E}}) }{\kappa^2_{+\mathrm{E}} - \kappa'^2_{+\mathrm{E}}} \nonumber\\
	& - 2 (L_{+\mathrm{E}}\cdot P_{\mathrm{E}})\, \mathbb{T}_{0\mathrm{E}} (\Pi'_{\mathrm{E}}\cdot \Pi_{\mathrm{E}}, \Pi'_{\mathrm{E}}\cdot \kappa_{\mathrm{E}}, \kappa'_{\mathrm{E}}\cdot \Pi_{\mathrm{E}}, \kappa'_{\mathrm{E}}\cdot \kappa_{\mathrm{E}}) \nonumber\\
	& \times \dfrac{B^+_{\mathrm{E}}(\kappa^2_{+\mathrm{E}}) - B^{+}_{\mathrm{E}}(\kappa'^2_{+\mathrm{E}}) }{\kappa^2_{+\mathrm{E}} - \kappa'^2_{+\mathrm{E}}} \bigg\} \, \mathbb{M}^+_{\mathrm{E}}(\kappa_{\mathrm{E}}^2, y, z) \mathbb{M}^-_{\mathrm{E}}(\kappa_{\mathrm{E}}^2, y, z) \nonumber\\
	& \times \mathbb{G}_{\mathrm{E}}(\kappa_{\mathrm{E}}^2,y,z) .
	\label{eq:emff_quark_Euc}
\end{align}
Within Eq.~\eqref{eq:emff_quark_Euc}, we have defined ${L_{+\mathrm{E}}^j = l_{\mathrm{E}}^j + \eta_{+} P_{\mathrm{E}}^j}$. Meanwhile, the momenta $\kappa$, $\kappa'$, $\Pi$, and $\Pi'$ are specified in Eq.~\eqref{eq:def_kappa_pm_kappap_pm}. The matrix $\mathbb{N}^+_{\mathrm{E}}{(\kappa'^2_{\mathrm{E}},\kappa'_{\mathrm{E}}\cdot \Pi'_{\mathrm{E}})}$ in Eq.~\eqref{eq:emff_quark_Euc} as the Euclidean-space version of Eq.~\eqref{eq:def_Np_bbm} is defined as
\begin{align*}
	&   \mathbb{N}^+_{\mathrm{E}}(\kappa'^2_{\mathrm{E}},\kappa'_{\mathrm{E}}\cdot \Pi'_{\mathrm{E}}) = \sigma^+_{\mathrm{sE}}(\kappa'^2_{+\mathrm{E}}) \mathbb{1}_{4} + \sigma_{\mathrm{vE}}^{+}(\kappa'^2_{+\mathrm{E}}) \nonumber\\ & \times 
	\left[ -k^{\mathrm{T}}_{\mathrm{R}}(-\kappa'^2_{\mathrm{E}}, -\kappa'_{\mathrm{E}}\cdot \Pi'_{\mathrm{E}}) - \eta_+ P^{\mathrm{T}}_{\mathrm{R}}(-\kappa'_{\mathrm{E}}\cdot \Pi'_{\mathrm{E}},M^2)\right],
\end{align*}
with $\sigma_{\mathrm{vE}}^+(\kappa'^2_{\mathrm{E}})$ and $\sigma_{\mathrm{sE}}^+(\kappa'^2_{\mathrm{E}})$ being the scalar components of the quark propagator given by Eq.~\eqref{eq:Euc_quark_prop_functions}. 

Following similar steps in deriving Eq.~\eqref{eq:emff_quark_Euc}, the EM form factor from the valence anti-quark in the Euclidean-space can be derived from Eq.~\eqref{eq:emff_antiquark_Minkowski}. Let us reuse Eq.~\eqref{eq:def_bbf_T0E} and Eq.~\eqref{eq:def_bbf_T1pmE} to represent the trace operations encountered. This EM from factor is then explicitly given by
\begin{align}
	& F_{-}(Q^2) = -\dfrac{e_{\bar{q}} N_{\mathrm{c}}}{2\pi^3 P^2_{\mathrm{E}}} \int_{0}^{+\infty}dl_{\mathrm{E}} \, l^3_{\mathrm{E}}\int_{-1}^{1}dy_{\mathrm{l}}\, \int_{-1}^{1}dz_{\mathrm{l}}\,\sqrt{1-z^2_{\mathrm{l}}} \nonumber\\
	& \times \overline{\mathbb{G}}_{\mathrm{E}}(\kappa'^2_{\mathrm{E}},y',z')
	\mathbb{N}^{-}_{\mathrm{E}}(\kappa'^2_{\mathrm{E}},\kappa'_{\mathrm{E}}\cdot \Pi'_{\mathrm{E}}) \bigg\{ \mathbb{T}_{1+\mathrm{E}} ( \Pi'_{\mathrm{E}}\cdot \Pi_{\mathrm{E}}, \Pi'_{\mathrm{E}}\cdot \kappa_{\mathrm{E}}, \nonumber\\
	& \kappa'_{\mathrm{E}}\cdot \Pi_{\mathrm{E}}, \kappa'_{\mathrm{E}}\cdot \kappa_{\mathrm{E}}, \Pi'_{\mathrm{E}}\cdot P_{\mathrm{E}}, \Pi_{\mathrm{E}}\cdot P_{\mathrm{E}}, \kappa'_{\mathrm{E}}\cdot P_{\mathrm{E}}, \kappa_{\mathrm{E}}\cdot P_{\mathrm{E}} ) \nonumber\\
	& \times \left[ \dfrac{A^-_{\mathrm{E}}(\kappa^2_{-\mathrm{E}}) + A^{-}_{\mathrm{E}}(\kappa'^2_{-\mathrm{E}}) }{2} + C_{\mathrm{MT},\mathrm{E}}( L^2_{-\mathrm{E}}, Q^2 ) \right] \nonumber \\
	& + 2 (L_{-\mathrm{E}}\cdot P_{\mathrm{E}})\,\mathbb{T}_{1+\mathrm{E}}
	( \Pi'_{\mathrm{E}}\cdot \Pi_{\mathrm{E}}, \Pi'_{\mathrm{E}}\cdot \kappa_{\mathrm{E}}, \kappa'_{\mathrm{E}}\cdot \Pi_{\mathrm{E}}, \kappa'_{\mathrm{E}}\cdot \kappa_{\mathrm{E}}, \nonumber\\
	& \Pi'_{\mathrm{E}}\cdot L_{-\mathrm{E}}, \Pi_{\mathrm{E}}\cdot L_{-\mathrm{E}}, \kappa'_{\mathrm{E}}\cdot L_{-\mathrm{E}}, \kappa_{\mathrm{E}}\cdot L_{-\mathrm{E}} ) \nonumber \\
	& \times
	\dfrac{A^-_{\mathrm{E}}(\kappa^2_{-\mathrm{E}}) - A^{-}_{\mathrm{E}}(\kappa'^2_{-\mathrm{E}}) }{\kappa^2_{-\mathrm{E}} - \kappa'^2_{-\mathrm{E}}} \nonumber \\
	& - 2 (L_{-\mathrm{E}}\cdot P_{\mathrm{E}} )\, \mathbb{T}_{0\mathrm{E}}(\Pi'_{\mathrm{E}}\cdot \Pi_{\mathrm{E}}, \Pi'_{\mathrm{E}}\cdot \kappa_{\mathrm{E}}, \kappa'_{\mathrm{E}}\cdot \Pi_{\mathrm{E}}, \kappa'_{\mathrm{E}}\cdot \kappa_{\mathrm{E}}) \nonumber \\
	& \times \dfrac{B^-_{\mathrm{E}}(\kappa^2_{-\mathrm{E}}) - B^{-}_{\mathrm{E}}(\kappa'^2_{-\mathrm{E}}) }{\kappa^2_{-\mathrm{E}} - \kappa'^2_{-\mathrm{E}}} \bigg\} \, \mathbb{M}^+_{\mathrm{E}}(\kappa_{\mathrm{E}}^2, y, z) \mathbb{M}^-_{\mathrm{E}}(\kappa_{\mathrm{E}}^2, y, z) \nonumber\\
	& \times 
	\mathbb{G}_{\mathrm{E}}(\kappa_{\mathrm{E}}^2,y,z), \label{eq:emff_antiquark_Euc}
\end{align}
where we have defined ${L_{-\mathrm{E}}^j = l_{\mathrm{E}}^j - \eta_{-} P_{\mathrm{E}}^j}$
and the Euclidean version of Eq.~\eqref{eq:def_bbm_n_antiquark} as
\begin{align*}
	&  \mathbb{N}^-_{\mathrm{E}}(\kappa'^2_{\mathrm{E}},\kappa'_{\mathrm{E}}\cdot \Pi'_{\mathrm{E}}) = \sigma^-_{\mathrm{sE}}(\kappa'^2_{-\mathrm{E}}) \mathbb{1}_{4} + \sigma_{\mathrm{vE}}^{-}(\kappa'^2_{-\mathrm{E}}) \nonumber\\
	& \times \left[ k^{\mathrm{T}}_{\mathrm{L}}(-\kappa'^2_{\mathrm{E}}, -\kappa'_{\mathrm{E}}\cdot \Pi'_{\mathrm{E}} ) - \eta_- P^{\mathrm{T}}_{\mathrm{L}}(-\kappa'_{\mathrm{E}}\cdot \Pi'_{\mathrm{E}}, M^2 ) \right] .
\end{align*}
Momenta $\kappa$, $\kappa'$, $\Pi$, and $\Pi'$ in Eq.~\eqref{eq:emff_antiquark_Euc} are defined according to Eq.~\eqref{eq:def_kappa_pm_kappap_pm}. 
\section{RESULTS\label{sc:results}}
\subsection{In the rest frame}
In order to solve for the pion structure numerically from its BSEs in Eq.~\eqref{eq:ps_BSE_Euc_rest_frame} and Eq.~\eqref{eq:ps_BSE_Euc_boosted_frame}, we apply the Maris-Tandy model where the gluon propagator becomes
\begin{align}
	& g^2\,\mathcal{G}_{\mathrm{E}}(k^2_{\mathrm{E}}) = \dfrac{4\pi^2}{\omega^6} d_{\mathrm{IR}}k^2_{\mathrm{E}}e^{-k^2_{\mathrm{E}}/\omega^2} \nonumber\\ & + \dfrac{8\pi^2\gamma_{m}}{ \ln\left[ e^2-1 + \left( 1 + k^2_{\mathrm{E}}/\Lambda^2_{\mathrm{QCD}} \right)^2 \right] } \dfrac{ 1-e^{-k^2_{\mathrm{E}}/(4m^2_t)} }{k^2_{\mathrm{E}}}, \label{eq:def_Maris_Tandy_model}
\end{align}
with $\gamma_m = 12 / ( 33 - 2 N_{\mathrm{f}} )$~\cite{Maris:1999nt}. The first and the second terms represent the interaction in the IR and the ultraviolet (UV). Parameters in Eq.~\eqref{eq:def_Maris_Tandy_model} are given in Table.~\ref{tab:pmts_Maris_Tandy_model}, resulting in the UV term identical to that in Ref.~\cite{Maris:1999nt}. The IR term is to be determined from bound state properties. 
\begin{table}
	\centering
	\begin{tabular}{ccccc}
		\hline
		$\omega$ & $d_{\mathrm{IR}}$ & $\Lambda_{\mathrm{QCD}}$ & $N_{\mathrm{f}}$ & $m_{t}$ \\
		\hline
		$0.4~\mathrm{GeV}$ & $0.859~(\mathrm{GeV})^2$ & $0.234~\mathrm{GeV}$ & $4$ & $0.5~\mathrm{GeV}$ \\
		\hline
	\end{tabular}
	\caption{Parameters of the gluon propagator in Eq.~\eqref{eq:def_Maris_Tandy_model}. The IR term is specified by its scale $\omega$ and strength $d_{\mathrm{IR}}$. While remaining parameters determine the UV term. } \label{tab:pmts_Maris_Tandy_model}
\end{table}

\begin{table}
	\centering
	\begin{tabular}{cccccc}
		\hline
		$m_{\mathrm{l}}$ & $\mu^2$ & $c$ & $n_{\mathrm{grid}}$ & $Z_2$ & $Z_{\mathrm{m}}$  \\
		\hline
		$3.6964~\mathrm{MeV}$ & $361.0~\mathrm{GeV}^2$ 
		& $10.0~\mathrm{GeV}^2$ & $2001$ &
		0.98201 & 0.67048 \\ 
		\hline
	\end{tabular}
	\caption{Parameters for the SDE of light-quark propagators. The renormalized quark mass $m_{\mathrm{l}}$ specifies the mass function at the renormalization scale $\mu^2$. The scale $c$ enters into Eq.~\eqref{eq:def_radial_map} with the resulting variable $u$ sampled by a grid of $n_{\mathrm{grid}}$ points applying the Gauss-Chebyshev quadrature of the first kind. Renormalization constants are given by $Z_2$ and $Z_m$. } \label{tab:pmts_SDE_light_quark}
\end{table}
\begin{figure*}
	\centering
	\includegraphics[width=0.75\linewidth]{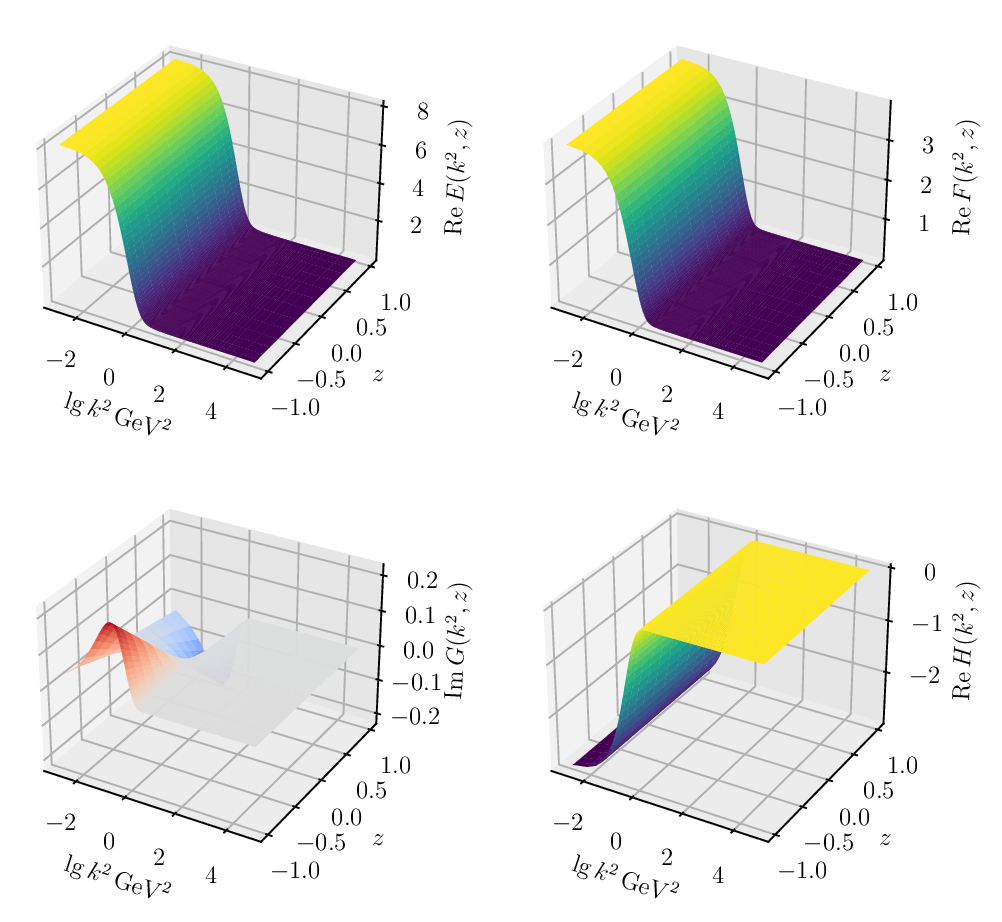}
	\caption{Scalar functions of the pion BSA in the rest frame. The imaginary parts of $E_{\mathrm{E}}{(k^2_{\mathrm{E}},z)}$, $F_{\mathrm{E}}{(k^2_{\mathrm{E}},z)}$, $H_{\mathrm{E}}{(k^2_{\mathrm{E}},z)}$, and the real part of $G_{\mathrm{E}}{(k^2_{\mathrm{E}},z)}$ are consistent with zero therefore not shown. }
	\label{fig:rfbsaquad}
\end{figure*}
With the gluon propagator and a given current quark mass, the SDE for the quark propagator can be solved numerically~\cite{Windisch:2016iud,Williams:2009wx}. In order to regularize the UV divergence in the self-energy, we impose $\Lambda^2_{\mathrm{UV}} = {1.0\times 10^6~\mathrm{GeV}^2}$ as the radial cutoff. The variable $k^2_{\mathrm{E}}$ in the quark propagators is mapped by
\begin{equation}
	u = k^2_{\mathrm{E}} / (k^2_{\mathrm{E}}+c) \label{eq:def_radial_map}
\end{equation}
with $c = 10.0~\mathrm{GeV}^2$ to the new variable $u$ located within ${[0,\,\Lambda^2_{\mathrm{UV}} / (\Lambda^2_{\mathrm{UV}} + c)} ]$. For integrals with respect to $k^2_{\mathrm{E}}$ via Eq.~\eqref{eq:def_radial_map} in the quark self-energy, we apply the Gauss-Chebyshev quadrature of the first kind with $2001$ points~\cite{abramowitz1964handbook,Jia:2024etj}. Such a rule is chosen because these integrals are end-point sensitive. After choosing the renormalization scale by $\mu = 19.0~\mathrm{GeV}$, the SDE for the quark propagator is solved by iteration to obtain the propagator in the spacelike region, the wave function renormalization $Z_2$, and the mass renormalization $Z_{\mathrm{m}}$~\cite{Jia:2024etj}. 

The parameters for the SDE of the light quark propagator are given by Table.~\ref{tab:pmts_SDE_light_quark}. We work in the isospin symmetric limit where the up and the down quarks are degenerate. 
The quark propagator at complex momenta is obtained from its SDE based on the spacelike self-energy~\cite{Jia:2024etj}. Within the kinematic region where the IR term in Eq.~\eqref{eq:def_Maris_Tandy_model} could not be densely sampled by the quadrature grid of the mapped variable, the Gauss-Hermite quadrature~\cite{abramowitz1964handbook} is alternatively applied. 

After solving for the propagators, the BSA for the pion in the rest-frame is solved from Eq.~\eqref{eq:ps_BSE_Euc_rest_frame} with subsequent normalization by Eq.~\eqref{eq:BSA_normalization_rest_frame}. Similar to the quark propagator, we map the radial momenta in the BSE applying Eq.~\eqref{eq:def_radial_map} with the same scale variable $c$ in the SDE. Because there is no divergence associated with these radial integrals, the Gauss-Chebyshev quadrature of the second kind is applied~\cite{abramowitz1964handbook,Jia:2024etj}. Specifically we construct $n_u = 121$ points for the grid of the variable $u\in[0.0, \,1.0]$. While a grid with $n_z = 15$ points is constructed for the angular variable $z\in[-1.0,\,1.0]$. 
\begin{table}
	\centering
	\begin{tabular}{cccccc}
		\hline
		$M (m_{\pi})$ & $c$ & $n_{u}$ & $n_{z}$ & $f_{\pi}$ & $r_{\pi}$ \\
		\hline
		$137.24~\mathrm{MeV}$ & $10.0~\mathrm{GeV}^2$ & $121$ & $15$ & $92.22~\mathrm{MeV}$ & $0.2355 ~\mathrm{GeV}^2$ \\
		\hline
	\end{tabular}
	\caption{Parameters and observables for the BSE of the pion in the rest frame. The pion mass enters as $M$ in Eq.~\eqref{eq:ps_BSE_Euc_rest_frame}. Radial variables are mapped by Eq.~\eqref{eq:def_radial_map} with scale $c$. The integers $n_{u}$ and $n_{z}$ specify the numbers of grid points for $u$ and $z$, respectively, both with Gauss-Chebyshev quadrature of the second type. The decay constant and the vertex residue obtained from the pion BSA are given by $f_{\pi}$ and $r_{\pi}$. } \label{tab:pmts_BSE_rest_frame} 
\end{table}

After choosing the grid for variables of the rest-frame BSA as in Table~\ref{tab:pmts_BSE_rest_frame}, the quadrature rules for the radial and the temporal-angle integrals in Eq.~\eqref{eq:ps_BSE_Euc_rest_frame} are known. The integral with respect to $y'$ involves the gluon propagator and the matrix ${\hat{\mathbb{T}}(k^2_{\mathrm{E}}, z, k'^2_{\mathrm{E}}, z', y')}$, both of which contain only known functions. We therefore adopt the adaptive Gauss-Kronrod quadrature~\cite{Galssi:GSL,Jia:2024etj} for its computation. Equation~\eqref{eq:ps_BSE_Euc_rest_frame} then becomes a linear operator on the scalar functions of the BSA. Similarly, the normalization condition in Eq.~\eqref{eq:BSA_normalization_rest_frame} takes the form of a bilinear operator. While for the derivative in Eq.~\eqref{eq:BSA_normalization_rest_frame}, we apply the following $3$-point finite difference approximation:
\begin{align}
	f'(M^2) & = \big[ 3 f(M^2) - 4 f(M^2-\Delta x) \nonumber\\
	& \quad + f(M^2-2\Delta x) \big] / (2\Delta x) + \mathcal{O}\left( \left(\Delta x\right)^2 \right) \label{eq:finite_difference_neg_2}
\end{align}
with $\Delta x = M^2 / 10$. Here only subtractions with respect to $M^2$ are taken to avoid sampling the quark propagator beyond what is specified in Eq.~\eqref{eq:quark_Euc_momentum_rest_frame}. With the pion BSA in the rest frame properly normalized, Eq.~\eqref{eq:fpi_Euc_rest_frame} and Eq.~\eqref{eq:rpi_Euc_rest_frame} yield the decay constant and the vertex residue. The resulting scalar functions in the Euclidean space are illustrated in Fig.~\ref{fig:rfbsaquad}. 

In the rest-frame computation, we choose a typical IR scale of $\omega = 0.4~\mathrm{GeV}$ in Eq.~\eqref{eq:def_Maris_Tandy_model}. For a given strength of the IR term, the light-quark mass is adjusted such that the ground state mass becomes the physical pion mass. We adjust the IR strength in conjunction to reproduce the experimental decay constant. The vertex residue is subsequently computed using Eq.~\eqref{eq:rpi_Euc_rest_frame}. These observables of the pion are given in Table~\ref{tab:pmts_BSE_rest_frame}, resulting in the generalized Gell-Mann--Oakes--Renner relation violated by $0.12\%$ in terms of the relative difference between two sides of Eq.~\eqref{eq:def_generalized_GMOR}. 
\subsection{The electromagnetic form factor of the pion}
\begin{table}
	\centering
	\begin{tabular}{cccc}
		\hline
		$c$ & $n_{u}$ & $n_{y}$ & $n_{z}$ \\
		\hline
		$0.5~\mathrm{GeV}^2$ & $41$ & $20$ & $15$ \\
		\hline
	\end{tabular}
	\caption{Parameters that determine the grid for scalar functions of the BSA in the boost frame. The scale variable $c$ enters into Eq.~\eqref{eq:def_radial_map} for the mapping of radial variables to $u\in[0.0,\,1.0]$. The integers $n_{u}$, $n_{y}$, and $n_{z}$ specify the numbers of grid points for $u$, $y$, and $z$, respectively. The Gauss-Chebyshev quadrature rule of the second type is applied. } \label{tab:pmts_boosted_frame}
\end{table}
After fixing model parameters in the rest-frame, we proceed to solve the BSE in the boost frames, where the quark propagators are sampled in the complex-momentum plane by the BSE in Eq.~\eqref{eq:ps_BSE_Euc_boosted_frame} within the region specified by Eq.~\eqref{eq:quark_Euc_momentum_boosted_frame}. The radial variable of the scalar functions in the BSA is mapped to $u$ by Eq.~\eqref{eq:def_radial_map} with $c = 0.5\,\mathrm{GeV}^2$. This mapped variable locates within ${[0.0,1.0]}$ because cutoffs are not required. While both of the angular variables are sampled within $[-1.0,1.0]$. The grid points for variables $u$, $y$, and $z$ are specified by the Gauss-Chebyshev rule of the second kind~\cite{abramowitz1964handbook,Jia:2024etj} with ${n_u = 41}$, ${n_y = 20}$, and ${n_z = 15}$, as in Table~\ref{tab:pmts_boosted_frame}. Since the $\phi$ integral in Eq.~\eqref{eq:ps_BSE_Euc_boosted_frame} only involves the gluon propagator and the known matrix ${\hat{\mathbb{T}}(k^2_{\mathrm{E}},y,z,k'^2_{\mathrm{E}},y',z',\phi')}$, we apply the adaptive Gauss-Kronrod quadrature for its computation~\cite{Galssi:GSL}. The quadrature rules for the remaining integrals in Eq.~\eqref{eq:ps_BSE_Euc_boosted_frame} are indicated by our choice of the correspond grids. The same quadrature rule also applies to the normalization condition in Eq.~\eqref{eq:BSA_normalization_boosted_frame}, with the derivative approximated by finite differences in Eq.~\eqref{eq:finite_difference_neg_2}. Similar numerical settings also apply to compute the decay constant and the vertex residue through Eq.~\eqref{eq:fpi_Euc_boosted_frame} and Eq.~\eqref{eq:rpi_Euc_boosted_frame}. 

At a given photon momentum, the EM form factors in the Euclidean space in Eq.~\eqref{eq:emff_quark_Euc} and Eq.~\eqref{eq:emff_antiquark_Euc} both take $2$ BSAs each carrying half of this momentum. We solve the corresponding BSAs from Eq.~\eqref{eq:ps_BSE_Euc_boosted_frame} with the bound state mass given by that in Table~\ref{tab:pmts_BSE_rest_frame}. The EM form factor then samples quark propagators within the same region by the boost-frame BSE. To evaluate the corresponding loop integrals in the EM form factor numerically, we apply the quadrature rule indicated by our choice of the grid in Table~\ref{tab:pmts_boosted_frame}. The radial integrals require no UV cutoff. And the integral with respect to the azimuthal angle is trivial. This choice of grid for the loop integrals samples the scalar functions of the BSAs at their native variables through the mapping in Eq.~\eqref{eq:varable_mapping_l_to_kappa}. Here linear interpolations are applied for these scalar functions. When ${\vert \kappa^2_{\mathrm{E}} - \kappa'^2_{\mathrm{E}}\vert \leq \Delta x}$, the $5$-point central finite difference rule is applied to approximate $[A_{\mathrm{E}}(\kappa^2_{\mathrm{E}}) - A_{\mathrm{E}}(\kappa'^2_{\mathrm{E}})] / ( \kappa^2_{\mathrm{E}} - \kappa'^2_{\mathrm{E}})$ and $[B_{\mathrm{E}}(\kappa^2_{\mathrm{E}}) - B_{\mathrm{E}}(\kappa'^2_{\mathrm{E}})] / ( \kappa^2_{\mathrm{E}} - \kappa'^2_{\mathrm{E}})$ through derivatives by
\begin{align}
	& f'(\kappa_{\mathrm{E}}^2) = \big[ f(\kappa_{\mathrm{E}}^2 - 2\Delta x) - 8 f(\kappa_{\mathrm{E}}^2 - \Delta x) + 8 f(\kappa_{\mathrm{E}}^2 + \Delta x) \nonumber\\
	& \hspace{1.5cm} - f(\kappa_{\mathrm{E}}^2 + 2\Delta x)\big] / ( 12 \Delta x)  + \mathcal{O}\left( \left(\Delta x\right)^4 \right)
\end{align}
with ${\Delta x = 1.0\,\mathrm{MeV}^2}$. 

When the photon momentum is sampled uniformly in the region of $\sqrt{Q^2}\in[0,\,Q_{\mathrm{max}}]$ with $N_{Q}$ number of points, a total number of $2N_{Q}-1$ BSAs are required to compute the EM form factor. This leads to the grid of the photon momentum in Table~\ref{tab:pion_emff}. While solving the boost-frame BSE in Eq.~\eqref{eq:ps_BSE_Euc_boosted_frame}, we adopt the bound state mass from the rest frame while allowing the eigenvalue to slightly deviate from unity due to differences in numerical accuracy. After obtaining this set of BSAs in the boost frame, we then compute the decay constant by Eq.~\eqref{eq:fpi_Euc_boosted_frame} and the vertex residue by Eq.~\eqref{eq:rpi_Euc_boosted_frame}. This results in the decay constant of the pion with zero spacial momentum in the boost frame to be $92.37\,\mathrm{MeV}$. Since the accurate determination of normalization for the BSA by Eq.~\eqref{eq:BSA_normalization_boosted_frame} is numerically difficult when the spacial momentum is large, we rescale the BSAs in the boost frame such that the decay constant becomes $92.37\,\mathrm{MeV}$. This rescaling factor then givens the relative error for the normalization of these BSAs. 
\begin{figure}
	\centering
	\includegraphics[width=\linewidth]{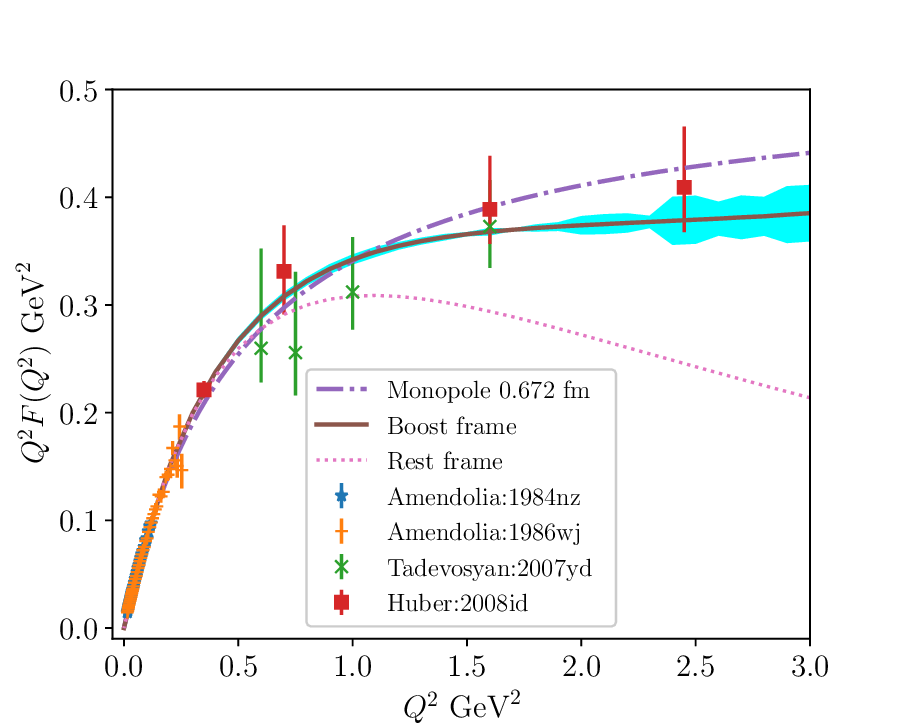}
	\caption{EM form factor of the pion applying the Ansatz vertex in Eq.~\eqref{eq:def_EM_vertex}. The solid line is the result from the boost-frame BSAs, with the error band corresponding to the uncertainty in the normalization. While the dotted line is the result from the rest-frame BSA. The dot-dash line corresponds to the monopole form in Eq.~\eqref{eq:emff_monopole} with an experimental charge radius of $0.672~\mathrm{fm}$. Experimental data points are based on Refs.~\cite{Amendolia:1984nz,NA7:1986vav,JeffersonLabFpi:2007vir,JeffersonLab:2008jve}. }
	\label{fig:pion_emff}
\end{figure}
\begin{figure}
	\centering
	\includegraphics[width=\linewidth]{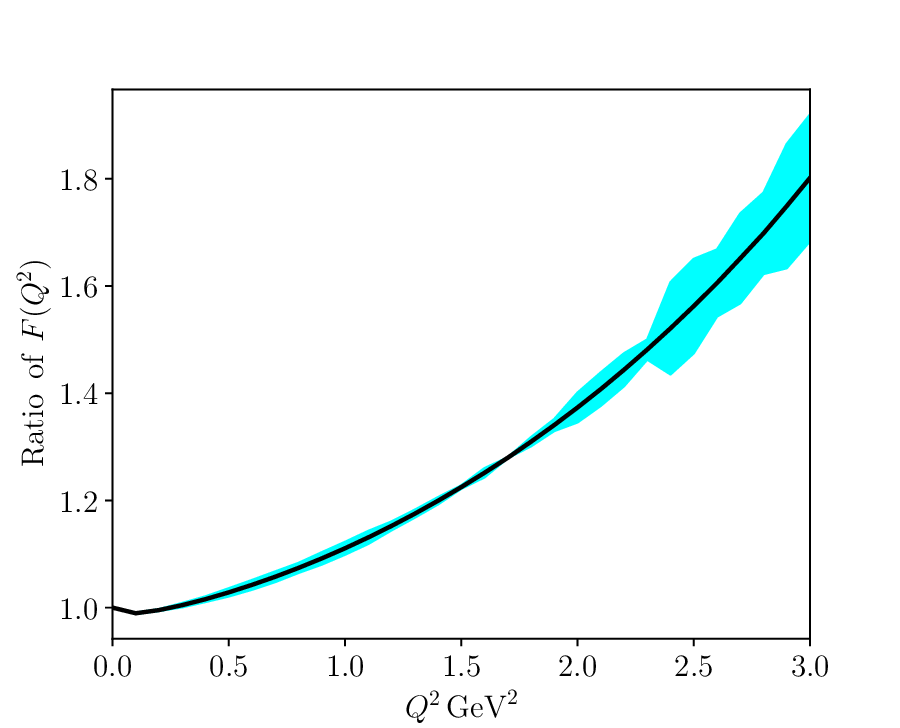}
	\caption{Ratio of the EM form factor for the pion from the boost-frame BSAs with that from the rest-frame BSA. The error band corresponds to the uncertainty in the normalization of the boost-frame BSA.}
	\label{fig:pion_emff_ratio}
\end{figure}

\begin{table}
	\centering
	\begin{tabular}{cccc}
		\hline
		$ Q^2\,\mathrm{GeV}^2 $ & $ F(Q^2) $ & $F(Q^2)$ error & $ F_{\mathrm{rf}}(Q^2) $ \\
		\hline
		0.0 & 1.000 & 0.000 & 1.000 \\
		0.1 & 0.860 & 0.001 & 0.869 \\
		0.2 & 0.753 & 0.001 & 0.756 \\
		0.3 & 0.666 & 0.003 & 0.663 \\
		0.4 & 0.594 & 0.003 & 0.585 \\
		0.5 & 0.534 & 0.004 & 0.519 \\
		0.6 & 0.483 & 0.004 & 0.464 \\
		0.7 & 0.440 & 0.004 & 0.416 \\
		0.8 & 0.402 & 0.003 & 0.375 \\
		0.9 & 0.370 & 0.004 & 0.339 \\
		1.0 & 0.342 & 0.004 & 0.308 \\
		1.1 & 0.318 & 0.003 & 0.281 \\
		1.2 & 0.296 & 0.002 & 0.257 \\
		1.3 & 0.276 & 0.002 & 0.235 \\
		1.4 & 0.259 & 0.001 & 0.216 \\
		1.5 & 0.244 & 0.001 & 0.199 \\
		1.6 & 0.230 & 0.002 & 0.184 \\
		1.7 & 0.218 & 0.000 & 0.170 \\
		1.8 & 0.206 & 0.001 & 0.158 \\
		1.9 & 0.196 & 0.002 & 0.146 \\
		2.0 & 0.187 & 0.004 & 0.136 \\
		2.1 & 0.179 & 0.004 & 0.127 \\
		2.2 & 0.171 & 0.004 & 0.118 \\
		2.3 & 0.164 & 0.002 & 0.111 \\
		2.4 & 0.158 & 0.009 & 0.104 \\
		2.5 & 0.152 & 0.009 & 0.097 \\
		2.6 & 0.146 & 0.006 & 0.091 \\
		2.7 & 0.141 & 0.007 & 0.086 \\
		2.8 & 0.137 & 0.006 & 0.080 \\
		2.9 & 0.132 & 0.009 & 0.076 \\
		3.0 & 0.128 & 0.008 & 0.071 \\
		\hline
	\end{tabular}
	\caption{EM form factors of the pion for a given momentum square of the photon. $F(Q^2)$ stands for the result from the BSAs in the boost frame. The error in $F(Q^2)$ is due to the uncertainty in the normalization of the BSAs. While $F_{\mathrm{rf}}(Q^2)$ corresponds to the result applying  the BSA with zero spacial momentum. } \label{tab:pion_emff}
\end{table}
In the computation of the EM form factor of the pion, we apply the quark-photon vertex given by the Ansatz in Eq.~\eqref{eq:def_EM_vertex}. Aside from using the BSAs in the boost frame, we also compute the EM form factor from only the BSA with zero spacial momentum. The results are given in Table~\ref{tab:pion_emff} with illustration in Fig.~\ref{fig:pion_emff}. The error for the EM form factor is based on that of the BSA normalization. The ratio of the EM form factor from the boost-frame BSAs with respect to that from the rest-frame is illustrated in Fig.~\ref{fig:pion_emff_ratio}. Because we work in the isospin symmetric limit, the contribution to the EM form factor from the valence quark is exactly twice of that from the valence anti-quark. The charge radius of the pion is obtained by fitting its EM form factor with the Ansatz vertex and boost-frame BSAs for ${Q^2\leq 3.0\,\mathrm{GeV}^2}$ by the following monopole form:
\begin{equation}
	F(Q^2) = 1 / \left( 1 + \langle r^2_{\pi} \rangle \, Q^2 / 6 \right) \label{eq:emff_monopole}
\end{equation}
Specifically, we obtain $\sqrt{\langle r^2_{\pi}\rangle} = 0.674\,\mathrm{fm}$. Our results of the EM form factor are different from those in Ref.~\cite{Maris:1999bh} because we did not apply the Chebyshev polynomial expansion for the angular dependence of the BSA, which also leads to different model parameters. 

As illustrated in Fig.~\ref{fig:pion_emff}, the difference in the EM form factors computed from the boost-frame BSAs and the rest-frame BSA is amplified while $Q^2$ increases. This is understood as the quarks interacting with the photon must carry the corresponding momenta. The lager these spacial momenta, the more difficult for these quarks to merge into the bound state in the rest frame. Therefore the EM form factor computed with the rest-frame BSA is smaller than that with the boost-frame BSAs. 
\section{SUMMARY\label{sc:summary}}
In the rainbow-ladder truncation, we converted the BSE for pseudoscalar bound states of a quark and an antiquark into coupled integral equations for the scalar components of the BSA. Similar decompositions were also carried out for the normalization of the BSA, the decay constant, and the vertex residue. The corresponding equations were subsequently obtained in the Euclidean space. 

Applying the Maris-Tandy modeling of the quark-gluon interaction, we solved the SDE for the propagators of light quarks~\cite{Jia:2024etj}. The BSE was then solved in the rest frame of the pion to determine its mass, decay constant, and vertex residue. With the strength for the IR term of the interaction and the light-quark mass both adjustable, we reproduced the experimental mass and decay constant of the pion.

To compute the EM form factor of the pion, we converted the impulse approximation expressions in terms of the scalar components of the BSA. These expressions sampled the BSAs for the bound state with corresponding spacial momenta. A matching set of these BSAs were therefore solved numerically from their BSE in the boost frame. The EM form factor was subsequently computed applying a phenomenological Ansatz for the quark-photon vertex. Comparisons were also made between this result and the one using only the BSA with zero spacial momentum. We extracted the charge radius of the pion by fitting a monopole form to the EM form factor. 
\begin{acknowledgments}
This work was supported by the US Department of Energy, Office of Science, Office of Nuclear Physics, under Contract No. DE-AC02-06CH11357. We gratefully acknowledge the computing resources provided on Bebop, a high-performance computing cluster operated by the Laboratory Computing Resource Center at Argonne National Laboratory.
\end{acknowledgments}
\appendix
\section{Quark Propagators Products in the BSE}
When the BSA is expanded into scalar functions by Eq.~\eqref{eq:ps_BSA_matrix}, its multiplications with the quark propagators are reduced into the left and right multiplications of $\slashed{P}$ and $\slashed{k}$ with the Dirac bases $T_j(k,P)$ in Eq.~\eqref{eq:def_Tj_basis}. Taking the left multiplication of $\slashed{P}$ with $T_j(k,P)$ as an example, it can be written in terms of matrix multiplications in the Dirac-basis representation. Specifically, we define the matries $P_{\mathrm{L},ij}(k\cdot P, P^2)$, $P_{\mathrm{R}}(k\cdot P, P^2)$, and $k{_\mathrm{L/R}}(k^2,k\cdot P)$ such that
\begin{subequations}
	\label{eq:multiplication_representation_bse}
	\begin{align}
		\slashed{P}\,T_j(k,P) &= \sum_{i=1}^{4}T_i(k,P)\,P_{\mathrm{L},ij}(k\cdot P, P^2), \\
		T_j(k,P)\,\slashed{P} &= \sum_{i=1}^{4}T_i(k,P)\,P_{\mathrm{R},ij}(k\cdot P, P^2), \\
		\slashed{k}\,T_j(k,P) &= \sum_{i=1}^{4}T_i(k,P)\,k_{\mathrm{L},ij}(k^2,k\cdot P), \\
		T_j(k,P)\,\slashed{k} &= \sum_{i=1}^{4}T_i(k,P)\,k_{\mathrm{R},ij}(k^2,k\cdot P),
	\end{align}
\end{subequations}
with $j\in\{1,\,2,\,3,\,4\}$ and the subscripts ``L'' and ``R'' specifying the direction of multiplications. These matrices are given by
\begin{subequations}
	\label{eq:multi_rule_LR}
	\begin{align}
		P_{\mathrm{L}}(k\cdot P, P^2) &= \begin{pmatrix}
			0 & P^2 & k\cdot P & 0 \\
			1 & 0 & 0 & k\cdot P \\
			0 & 0 & 0 & -P^2 \\
			0 & 0 & -1 & 0
		\end{pmatrix}, \\[0.5em]
		P_{\mathrm{R}}(k\cdot P, P^2) &= \begin{pmatrix}
			0 & P^2 & k\cdot P & 0 \\
			1 & 0 & 0 & -k\cdot P \\
			0 & 0 & 0 & P^2 \\
			0 & 0 & 1 & 0
		\end{pmatrix}, \\[0.5em]
		k_{\mathrm{L}}(k^2,k\cdot P) &= \begin{pmatrix}
			0 & k\cdot P & k^2 & 0 \\
			0 & 0 & 0 & k^2 \\
			1 & 0 & 0 & -k\cdot P \\
			0 & 1 & 0 & 0
		\end{pmatrix}, \\[0.5em]
		k_{\mathrm{R}}(k^2,k\cdot P) &= \begin{pmatrix}
			0 & k\cdot P & k^2 & 0 \\
			0 & 0 & 0 & -k^2 \\
			1 & 0 & 0 & k\cdot P \\
			0 & -1 & 0 & 0
		\end{pmatrix}. 
	\end{align}
\end{subequations}
We then obtain the multiplications of the quark propagator to the BSA in the basis representation based on $k^\mu_\pm = k^\mu\pm \eta_\pm P^\mu$. Defining the left and right multiplication matrices as
\begin{subequations} 
	\label{eq:def_matrix_bbM}
	\begin{align}
		& \mathbb{M}^+(k^2,k\cdot P) = -\sigma^+_{\mathrm{s}}(k^2_+) \mathbb{1}_{4} + \sigma^+_{\mathrm{v}}(k_+^2) \nonumber \\
		& \times 
		\left[ k_{\mathrm{L}}(k^2,k\cdot P) + \eta_+ P_{\mathrm{L}}(k\cdot P, P^2) \right] 
	\end{align}
	and 
	\begin{align}
		& \mathbb{M}^-(k^2,k\cdot P) = \sigma^-_{\mathrm{s}}(k^2_-)\mathbb{1}_{4} + \sigma^-_{\mathrm{v}}(k^2_-) \nonumber \\
		& \times \left[ k_{\mathrm{R}}(k^2,k\cdot P) - \eta_- P_{\mathrm{R}}(k\cdot P,P^2) \right]
	\end{align}
	gives 
	\begin{equation}
		\mathbb{M}(k^2,k\cdot P) = \mathbb{M}^+(k^2,k\cdot P)\,\mathbb{M}^-(k^2,k\cdot P).
	\end{equation}
\end{subequations}
The multiplications of the quark propagators then map the Dirac bases of the BSA in Eq.~\eqref{eq:def_Tj_basis} according to
\begin{equation}
	S(k_+) \gamma_5 T(k,P) S(k_-) = -\gamma_5 T(k,P) \mathbb{M}(k^2,k\cdot P). 
	\label{eq:apply_matrix_bbM}
\end{equation}
\section{Representation of the trace projection\label{app:trace_rep_bse}}
To calculate the trace projection of the BSE onto the Dirac bases in Eq.~\eqref{eq:def_Tj_basis}, let us start with the following trace of the Dirac bases 
\begin{align}
	& \mathrm{Tr}\,\gamma_{5}\,\overline{T}(k,P)\,\gamma_{5}\,T(k,P) \nonumber\\
	& = 4\begin{pmatrix}
		1 & 0         & 0         & 0 \\
		0 & -P^2      & -k\cdot P & 0 \\
		0 & -k\cdot P & -k^2      & 0 \\
		0 & 0         & 0         & (k\cdot P)^2 - k^2P^2
	\end{pmatrix}. \label{eq:Dirac_trace_lhs_BSE}
\end{align}
While applying Eq.~\eqref{eq:apply_matrix_bbM} to the right-hand side of BSE in Eq.~\eqref{eq:ps_BSE} gives
\begin{align}
	\Gamma(k,P) & = ig^2 C_{\mathrm{F}}\int \ddu \ell\
	\mathcal{G}(q^2)\ \gamma_5\gamma^\mu \, 
	T(\ell,P)\mathbb{M}(\ell^2,\ell\cdot P) \nonumber\\
	& \quad \times 
	\gamma^\nu\left(-g_{\mu\nu}+ q_\mu q_\nu / q^2 \right) 
	\mathbb{G}(\ell^2,\ell\cdot P). 
	\label{eq:ps_BSE_merged}
\end{align}
To represent the trace projections of the right-hand side of Eq.~\eqref{eq:ps_BSE_merged}, let us define the matrix ${\mathbb{T}(k^2,k\cdot P, \ell^2, \ell\cdot P, k\cdot \ell)}$ as
\begin{align}
	&\mathbb{T}(k^2,k\cdot P, \ell^2, \ell\cdot P, k\cdot \ell) 
	= (g_{\mu\nu}-q_{\mu} q_{\nu} / q^2) \nonumber\\
	& \times 
	\mathrm{Tr}\,\gamma_5\,\overline{T}(k,P) \,\gamma_5\gamma^\mu T(\ell, P) \gamma^\nu,
	\label{eq:def_matrix_bbT}
\end{align}
with $T(k,P)$ and $\overline{T}(k,P)$ given by Eq.~\eqref{eq:def_Tj_basis} and Eq.~\eqref{eq:def_Tj_basis_transpose}. Combining Eq.~\eqref{eq:Dirac_trace_lhs_BSE} and Eq.~\eqref{eq:def_matrix_bbT} results in 
\begin{align}
	&\hat{\mathbb{T}}(k^2,k\cdot P, \ell^2, \ell\cdot P, k\cdot \ell) \nonumber\\
	&= \frac{1}{4}
	\begin{pmatrix}
		1 &  0        &  0        & 0 \\
		0 & -P^2      & -k\cdot P & 0 \\
		0 & -k\cdot P & -k^2      & 0 \\
		0 &  0        &  0        & (k\cdot P)^2 - k^2P^2
	\end{pmatrix}^{-1} \nonumber\\
	& \times 
	\begin{pmatrix}
		12 & 0 & 0 & 0 \\
		0 & 4[P^2 + 2(P\cdot q)^2/q^2] & \begin{aligned} 4[\ell\cdot P + 2(P\cdot q) \\ \times (\ell\cdot q) / q^2] \end{aligned} & 0 \\
		0 & \begin{aligned} 4[k\cdot P + 2(k\cdot q) \\ \times(P\cdot q) / q^2] \end{aligned} & \begin{aligned} 4[k\cdot \ell + 2(k\cdot q) \\ \times (\ell\cdot q )/ q^2] \end{aligned} & 0 \\
		0 & 0 & 0 & \mathbb{T}_{44}
	\end{pmatrix}, \label{eq:def_That_bbm}
\end{align}
with $q = \ell -k$ and 
\begin{align}
	& \mathbb{T}_{44} = \dfrac{4}{q^2}\big\{ 2 \big[(k\cdot P)(\ell\cdot P)(k\cdot \ell) - (k\cdot P)^2 \ell^2 - k^2(\ell\cdot P)^2 \nonumber\\
	& \! + k^2P^2\ell^2 \big] + (k^2+\ell^2) \big[ (k\cdot P)(\ell\cdot P) - P^2(k\cdot \ell) \big] \big\}. 
\end{align}
The trace projection of the BSE in Eq.~\eqref{eq:ps_BSE_merged} then becomes Eq.~\eqref{eq:ps_BSE_projected}. 

In the rest frame, based on the definition in Eq.~\eqref{eq:def_TE_rest_frame} we have the following trace matrix in the Euclidean space
\begin{align}
	& \quad \hat{\mathbb{T}}_{\mathrm{E}}(k_{\mathrm{E}}^2, z, k'^2_{\mathrm{E}}, z',y') \nonumber\\& = 
	\begin{pmatrix}
		3 & 0 & 0 & 0 \\
		0 & -1 & -\dfrac{ik'_{\mathrm{E}}}{M}\left( z' - zy'\sqrt{\dfrac{1-z'^2}{1-z^2}} \right) & 0 \\
		0 & 0 & -\dfrac{k'_{\mathrm{E}}y'}{k_{\mathrm{E}}}\sqrt{\dfrac{1-z'^2}{1-z^2}} & 0 \\
		0 & 0 & 0 & 0
	\end{pmatrix}  \nonumber\\
	& \quad + \frac{2}{q^2_{\mathrm{E}}}
	\begin{pmatrix}
		0 & 0 & 0 & 0 \\
		0 & -\ell_{\mathrm{E}}c_1 r_1 & -\frac{i\ell^2_{\mathrm{E}}}{M}c_2 r_1 & 0 \\
		0 & -\frac{iM}{k_{\mathrm{E}}}c_1 r_2 & \frac{\ell_{\mathrm{E}}}{k_{\mathrm{E}}}c_2 r_2 & 0 \\
		0 & 0 & 0 & \hat{T}_{\mathrm{E}44}
	\end{pmatrix} ,\label{eq:Euclidean_bbm_trace_matrix}
\end{align}
with
\begin{subequations}
	\begin{align}
		c_1 &= \ell_{\mathrm{E}}z' - k_{\mathrm{E}} z, \\
		c_2 &= \ell_{\mathrm{E}}-\zeta(z,z',y') k_{\mathrm{E}}, \\
		r_1 &= z' - zy'\sqrt{(1-z'^2)/(1-z^2)}, \\
		r_2 &= k_{\mathrm{E}} - \ell_{\mathrm{E}}y'\sqrt{{1-z'^2}{1-z^2}}, \\
		\hat{T}_{\mathrm{E}44} &= \left[ \ell_\mathrm{E}(\ell^2_\mathrm{E} + k^2_{\mathrm{E}}) / ( 2k_{\mathrm{E}} ) - zz'\ell^2_{\mathrm{E}} \right] \nonumber\\
		& \quad\times y'\sqrt{(1-z'^2)/(1-z^2)} -\ell^2_{\mathrm{E}}\,(1 - z'^2).
	\end{align}
\end{subequations}
While in the boost frame, the definition in Eq.~\eqref{eq:def_TE_boosted_frame} results in
\begin{align}
	&\hat{\mathbb{T}}_{\mathrm{E}}(k^2_{\mathrm{E}},y,z,\ell^2_{\mathrm{E}},y',z',\phi) = 
	\begin{pmatrix}
		1 & 0 & 0 & 0 \\
		0 & -\dfrac{k^2_{\mathrm{E}}}{\Delta_{\mathrm{E}}} & \dfrac{k_{\mathrm{E}}\cdot P_{\mathrm{E}}}{\Delta_{\mathrm{E}} } & 0 \\[0.8em]
		0 & \dfrac{k_{\mathrm{E}}\cdot P_{\mathrm{E}}}{\Delta_{\mathrm{E}} } & \dfrac{M^2}{\Delta_{\mathrm{E}} } & 0 \\
		0 & 0 & 0 & \dfrac{1}{\Delta_{\mathrm{E}} }
	\end{pmatrix} \nonumber\\
	& \times
	\begin{pmatrix}
		3 & 0 & 0 & 0 \\
		0 & M^2 - \dfrac{2c_1^2}{q^2_{\mathrm{E}}} & -\ell_{\mathrm{E}}\cdot P_{\mathrm{E}} - \dfrac{2c_1r'}{q^2_{\mathrm{E}}} & 0 \\
		0 & -k_{\mathrm{E}}\cdot P_{\mathrm{E}} - \dfrac{2c_1r}{q^2_{\mathrm{E}}} & -k_{\mathrm{E}}\cdot \ell_{\mathrm{E}} - \dfrac{2rr'}{q^2_{\mathrm{E}}} & 0 \\
		0 & 0 & 0 & \tilde{\mathbb{T}}_{44\mathrm{E}}
	\end{pmatrix}, \label{eq:Euclidean_bbm_trace_matrix_boosted_frame}
\end{align}
with $\Delta_{\mathrm{E}} = (k_{\mathrm{E}}\cdot P_{\mathrm{E}})^2 + k_{\mathrm{E}}^2M^2$, 
$ c_1 = P_{\mathrm{E}}\cdot q_{\mathrm{E}} $, $r = k_{\mathrm{E}}\cdot q_{\mathrm{E}} $, $r' = \ell_{\mathrm{E}}\cdot q_{\mathrm{E}} $, 
and
\begin{align}
	\tilde{\mathbb{T}}_{44\mathrm{E}} & = \big\{ 2\big[(k_{\mathrm{E}}\cdot P_{\mathrm{E}})(\ell_{\mathrm{E}}\cdot P_{\mathrm{E}})(k_{\mathrm{E}}\cdot \ell_{\mathrm{E}}) - (k_{\mathrm{E}}\cdot P_{\mathrm{E}})^2\ell^2_{\mathrm{E}} \nonumber\\ 
	& \quad -k^2_{\mathrm{E}}(\ell_{\mathrm{E}}\cdot P_{\mathrm{E}})^2 - k^2_{\mathrm{E}}\ell^2_{\mathrm{E}}M^2 \big] + (k^2_{\mathrm{E}} + \ell^2_{\mathrm{E}}) \nonumber \\
	& \quad \times \big[ (k_{\mathrm{E}}\cdot P_{\mathrm{E}})(\ell_{\mathrm{E}}\cdot P_{\mathrm{E}}) + (k_{\mathrm{E}}\cdot \ell_{\mathrm{E}})M^2 \big] \big\} / q^2_{\mathrm{E}}.
\end{align}
\section{Quark Propagator Products in Normalization Condition\label{sc:trace_operations}}
In order to convert the normalization condition of the BSE in terms of its scalar components, we need to calculate the mapping of the Dirac bases by the multiplications of quark propagators. Because the momentum $Q$ in the quark propagators does not match the bound state momentum $P$ in Eq.~\eqref{eq:BSA_normalization}, the Dirac bases is expanded from Eq.~\eqref{eq:def_Tj_basis} into 
\begin{align}
	T(k,P,Q) & = \Big( \mathbb{1},\, \slashed{P},\, \slashed{k},\, -i\sigma_{kP},\, \slashed{Q},\, -i\sigma_{PQ},\, -i\sigma_{kQ},\, \nonumber\\
	& \quad \{\slashed{Q},\,-i\sigma_{kP}\}/2 \Big). \label{eq:extended_basis}
\end{align}
Similar to Eq.~\eqref{eq:multiplication_representation_bse}, let us define the matrices $I_{\mathrm{L/R}}$, $K_{\mathrm{L/R}}$, and $Q_{\mathrm{L/R}}$ according to
\begin{subequations}
	\label{eq:multiplication_representation_normalization}
	\begin{align}
		T_j(k,P,Q)\mathbb{1} & = \sum_{i=1}^{8}T_i(k,P,Q) I_{\mathrm{R},ij}, \\
		T_j(k,P,Q)\slashed{k} & = \sum_{i=1}^{8}T_i(k,P,Q)K_{\mathrm{R},ij}(k^2,k\cdot P), \\
		T_j(k,P,Q)\slashed{Q} & = \sum_{i=1}^{8}T_i(k,P,Q)Q_{\mathrm{R},ij}(k\cdot Q, P\cdot Q),
	\end{align}
\end{subequations}
for $j\in\{1,\,2,\,3,\,4\}$, and
\begin{subequations}
	\begin{align}
		\mathbb{1}T_j(k,P,Q) & = \sum_{i=1}^{8}T_i(k,P,Q) I_{\mathrm{L},ij}, \\
		\slashed{k}T_j(k,P,Q) & = \sum_{i=1}^{8}T_i(k,P,Q)K_{\mathrm{L},ij}(k^2,k\cdot P,k\cdot Q), \\
		\slashed{Q}T_j(k,P,Q) & = \sum_{i=1}^{8}T_i(k,P,Q)Q_{\mathrm{L},ij}(k\cdot Q, P\cdot Q, Q^2), 
	\end{align}
\end{subequations}
for $j\in\{1,\,2,\,3,\,4,\,5,\,6,\,7,\,8\}$. Since we choose to first calculate the right multiplication of the quark propagator to the BSA, $I_{\mathrm{R}}$, ${K_{\mathrm{R}}(k^2,k\cdot P)}$, and ${Q_{\mathrm{R}}(k\cdot Q, P\cdot Q)}$ are $8$-by-$4$ matrices. While the remaining ones in Eq.~\eqref{eq:multiplication_representation_normalization} are $8$-by-$8$ matrices. They are explicitly given by
\begin{subequations}
	\label{eq:def_matrices_normalization}
	\begin{equation}
		I_{\mathrm{R}} =
		\begin{pmatrix}
			\mathbb{1}_{4} \\
			\mathbb{0}_{4}
		\end{pmatrix}, 
	\end{equation}
	\begin{equation}
		K_{\mathrm{R}}(k^2,k\cdot P) = 
		\begin{pmatrix}
			k_{\mathrm{R}}(k^2,k\cdot P) \\
			\mathbb{0}_{4}
		\end{pmatrix}, 
	\end{equation}
	\begin{equation}
		Q_{\mathrm{R}}(k\cdot Q,P\cdot Q) = 
		\begin{pmatrix}
			R(k\cdot Q, P\cdot Q ) \\
			\mathbb{1}_{4}
		\end{pmatrix}, 
	\end{equation}
\end{subequations}
with
\begin{equation}
	R(k\cdot Q, P\cdot Q ) = 
	\begin{pmatrix}
		0 & P\cdot Q & k\cdot Q & 0\\
		0 & 0 & 0 & -k\cdot Q \\
		0 & 0 & 0 & P\cdot Q \\
		0 & 0 & 0 & 0 
	\end{pmatrix} ,
\end{equation} 
$\mathbb{1}_4= {\mathrm{diag}\{1,\,1,\,1,\,1\}}$, and $\mathbb{0}_{4} = {\mathrm{diag}\{0,\,0,\,0,\,0\}}$ for the right multiplications. We also obtain
${I_{\mathrm{L}}} = \mathrm{diag}\,\{1,\, 1,\, 1,\, 1,\, 1,\, 1,\, 1,\, 1\}$, 
\begin{align}
	K_{\mathrm{L}}(k^2,k\cdot P, k\cdot Q) = 
	\begin{pmatrix}
		k_{\mathrm{L}}(k^2,k\cdot P) & k\cdot Q\,J_{\mathrm{L}} \\
		\mathbb{0}_{4} & k_{\mathrm{L}}(k^2,k\cdot P)
	\end{pmatrix}, \\
	Q_{\mathrm{L}}(k\cdot Q, P\cdot Q, Q^2) =
	\begin{pmatrix}
		R(k\cdot Q,P\cdot Q) & Q^2\,J_{\mathrm{L}} \\
		J_{\mathrm{L}} & R(k\cdot Q, P\cdot Q)
	\end{pmatrix},
\end{align}
with $J_{\mathrm{L}} = {\mathrm{diag}\{1,-1,-1,1\}}$ for the left multiplications in Eq.~\eqref{eq:multiplication_representation_normalization}. 
\section{THE NORMALIZATION $N_\rho$\label{sc:n_rho}}
The Ansatz transverse vertex in Eq.~\eqref{eq:def_EM_vertex} follows from a model for the $\rho$ meson BSA~\cite{Maris:1999bh} given by
\begin{equation}
	\Gamma^\mu_\rho(q,P) = \left(\gamma^\mu - \dfrac{P^\mu \sh{P}}{P^2} \right) 
	\frac{N_{\rho}}{1 + \left( q^2/\omega^2 \right)^2}. 
	\label{eq:Ansatz_rho_BSA}
\end{equation}
The leptonic decay constant for the $\rho$ meson reads~\cite{Maris:1999nt}
\begin{equation}
	f_\rho\,m_\rho = -iZ_2 \, \mathrm{Tr}\int \ddu q\ 
	\gamma_\mu\, S^+(q_+)\Gamma^\mu_\rho(q,P)S^-(q_-), 
	\label{eq:def_f_rho}
\end{equation}
with $q_\pm = q \pm \eta_{\pm}P$. Substituting Eq.~\eqref{eq:Ansatz_rho_BSA} into Eq.~\eqref{eq:def_f_rho} gives 
\begin{align}
	& f_\rho m_\rho = \frac{Z_2N_\rho}{\pi^3}\int_{0}^{\infty} \dd q_{\mathrm{E}}\,  q_{\mathrm{E}}^3
	\int_{-1}^{1}\dd z\, 
	\frac{\sqrt{1-z^2}}{1 + q^4_{\mathrm{E}}/\omega^4} \nonumber\\
	& \times \!
	\Big\{3\sigma_{\mathrm{sE}}^+(q_{+\mathrm{E}}^2) \sigma_{\mathrm{sE}}^-(q_{-\mathrm{E}}^2) + \sigma_{\mathrm{vE}}^+(q_{+\mathrm{E}}^2) \sigma_{\mathrm{vE}}^-(q_{-\mathrm{E}}^2) \big[ q_{\mathrm{E}}^2(1+2z^2)  \nonumber\\
	& \quad + 3\big(\eta_+\eta_-m^2_\rho- i(\eta_+-\eta_-)m_\rho q_{\mathrm{E}} z\big) 
	\big] \Big\},
	\label{eq:rho_decay_constant_Euc}
\end{align}
in the Euclidean space with $ {q^2_{\pm\mathrm{E}}} = {q_{\mathrm{E}}^2} \mp {2i\eta_{\pm}m_{\rho}q_{\mathrm{E}}z} - {\eta^2_\pm m^2_{\rho}} $
and $\eta_{\pm} = 1/2$. Here $\sigma^{\pm}_{\mathrm{vE}}(q^2_{\pm\mathrm{E}})$ and $\sigma^{\pm}_{\mathrm{sE}}(q^2_{\pm\mathrm{E}})$ are the scalar components of the quark propagator defined by Eq.~\eqref{eq:Euc_quark_prop_functions}. They are solved from the SDE for propagators of light quarks with parameters in Table~\ref{tab:pmts_Maris_Tandy_model} and Table~\ref{tab:pmts_SDE_light_quark}~\cite{Jia:2024etj}. The normalization constant $N_\rho$ is determined from Eq.~\eqref{eq:rho_decay_constant_Euc} given the mass and the decay constant of the $\rho$ meson in Table~\ref{tab:pmts_MTA_vertex}.
\bibliography{BSE_pion_bib.bib}
\end{document}